\pdfoutput=1
\documentclass[twocolumn,epjc3]{svjour3}

\RequirePackage[T1]{}

\usepackage{amsmath}
\usepackage{enumitem}

\smartqed

\RequirePackage{graphicx}
\RequirePackage{mathptmx}
\RequirePackage{flushend}
\usepackage{multirow}
\usepackage{amssymb}
\usepackage{amsmath}
\usepackage{romannum}
\RequirePackage[numbers,sort&compress]{natbib}
\RequirePackage[colorlinks,citecolor=blue,urlcolor=blue,linkcolor=blue]{hyperref}

\DeclareMathOperator{\sech}{sech}

\journalname{}

\begin{document}
	
	\title{The Einstein Telescope standard siren simulations for $f(Q)$ cosmologies}

	\author{
		Xianfu Su\thanksref{addr1}
		\and
		Dongze He\thanksref{addr1}
		and
		Yi Zhang\thanksref{e1,addr1}}
	
	\thankstext{e1}{e-mail: zhangyia@cqupt.edu.cn}
	\institute{Chongqing University
		of Posts and Telecommunications, Chongqing 400065, China\label{addr1}}
	
	\date{Received: date / Accepted: date}

	\maketitle


\begin{abstract}
To investigate the model and extra frictional effects in standard siren simulation of $f(Q)$ cosmologies, we simulated  three types of standard siren data based on different fiducial models ($\Lambda$CDM and $f(Q)$ models).
Both effects are important in standard siren simulation. Explicitly, the $f(Q)_P$ and $f(Q)_E$ models need more observational data (e.g.growth factor) to further study. The $f(Q)_{PE}$ model could be ruled out by the EM data. And both the $f(Q)_{HT}$ models will be excluded by the future standard siren data.
\end{abstract}
\section{Introduction}
The  standard siren (SS) of gravitational wave (GW) provides an absolute  measurement of distance without dependence on other sources \cite{Schutz:1986gp}. This standard siren method is widely used to constrain cosmological models  especially for the modified gravity. 
Presently, the direct detection of gravitational wave has  discovered at least 99 standard siren events \cite{LIGOScientific:2016aoc,LIGOScientific:2016dsl,LIGOScientific:2017adf,LIGOScientific:2017bnn,LIGOScientific:2017ycc,LIGOScientific:2020ibl,LIGOScientific:2021qlt}, but only one single confirmed standard siren event (GW170817) and one possible standard siren event (GW1905
21) have been detected. These two events  are unable to do effective cosmological constraints. 
In the coming decade, ground-based (e.g.Einstein Telescope (ET)  \cite{Punturo:2010zz,Sathyaprakash:2009xt,Zhang:2018byx} and space-based telescopes (e.g.Taiji \cite{Hu:2017mde}, Tianqin \cite{TianQin:2015yph}, and LISA \cite{Lisa}) experiments are predicted to discover more standard siren sources. Therefore, to forecast fundamental properties of gravity, the mock catalogs of standard sirens should be created.

In the standard siren method, the luminosity distance $D_{L}$  could be  extracted  from the GW amplitude $h_{A}$. And the standard siren simulation should be based on the background cosmology  but it could  not be determined at present. 
The cosmological constant called $\Lambda$CDM model is the simplest theoretical explanation for our accelerating universe which is preferred by the majority of observational survey releases (e.g.Planck data \cite{Planck:2015fie,Planck:2013pxb}).
And, dark energy  and modified gravity are the candidates of explaining the accelerating universe  as well. All the cosmological models could 
affect the amplitude of standard siren. Then conversely, the standard siren data could constrain cosmological models.  
Here, the choice of  fiducial model of standard siren simulation is called  the model effect.
And compared with the $\Lambda$CDM and dark energy models of general relativity (GR), the propagation equation of  gravitational wave in modified gravity has  an  extra friction term which affects the standard siren simulation as well. It  is called the extra frictional effect. Roughly, based on affine connections \cite{Braglia:2020auw,Brax:2013fda,Clifton:2011jh,Lin:2018nxe,DiValentino:2015bja,Lee:2022gzh}, there are mainly three types of modified gravity: $f(R)$, $f(T)$ and $f(Q)$ cosmologies.     
Here, we choose  the $\Lambda$CDM model in general relativity (GR) as baseline and discuss the $f(Q)$ cosmologies which are relatively simple. Among the various $f(Q)$ models \cite{Bengochea:2008gz,Chen:2010va,Linder:2010py,Anagnostopoulos:2021ydo,Ferreira:2023awf,Wu:2010av,Anagnostopoulos:2022gej,Qi:2017xzl,BeltranJimenez:2017tkd,BeltranJimenez:2019tme,Heisenberg:2023lru,Lazkoz:2019sjl,Ayuso:2020dcu,Mandal:2021bpd,Barros:2020bgg,Atayde:2021pgb,Ferreira:2023tat,Nojiri:2024hau,Nojiri:2024zab,Bajardi:2023vcc,Frusciante:2021sio,Paliathanasis:2023nkb,Shabani:2023nvm,Sokoliuk:2023ccw,Flathmann:2020zyj,Solanki:2022ccf,Capozziello:2022wgl,Narawade:2022jeg,Dimakis:2022rkd,Myrzakulov:2023ohc,Pati:2022dwl,Khyllep:2022spx,Koussour:2023ulc,Oliveros:2023mwl,Albuquerque:2022eac,Atayde:2023aoj,
Goncalves:2024sem,Heisenberg:2023wgk,Bahamonde:2021gfp,Sakr:2024eee,Capozziello:2022zzh,Capozziello:2023vne,Xu:2018npu,Briffa:2021nxg}, two $\Lambda$CDM-like models (the power-law $f(Q)_{P}$   \cite{Bengochea:2008gz,Chen:2010va} and  square-root exponential $f(Q)_{E}$ \cite{Linder:2010py} models) which could come back to $\Lambda$CDM model are chosen to discuss. Correspondingly, two non $\Lambda$CDM-like models (the power exponential $f(Q)_{PE}$    \cite{Anagnostopoulos:2021ydo,Ferreira:2023awf} and  hyperbolic tangent $f(Q)_{HT}$  \cite{Wu:2010av,Anagnostopoulos:2022gej,Qi:2017xzl} models) are chosen to constrain as well.
\footnote{
	The description of gravity using  Teleparallel Gravity with Weitaenb$\ddot{o}$ck connection $T$ which is  called  torsion scalar is equivalent with the Symmetric Teleparallel  gravity using the non-metricity scalar $Q$ in the background level. In the both gravities, the curvature $R$ in GR is replaced by $T$ or $Q$ \cite{Ferraro:2012wp,Zhang:2011qp,Awad:2017yod,Bamba:2010iw,Paliathanasis:2016vsw,Bamba:2010wb,Nassur:2016gzr,Salako:2013gka,Nesseris:2013jea,	Nunes:2016qyp,Nunes:2018xbm,Basilakos:2018arq,Zhang:2012jsa,Capozziello:2018hly,Nunes:2018evm,Nunes:2019bjq,Yan:2019gbw,Jensko:2024bee}.} 

To invest the mode and extra friction effects,  we intend to simulate three types of mock standard siren data  in this paper: the first  one (SS\Romannum{1}$_\Lambda$) is based on the $\Lambda$CDM model; the second one  (SS\Romannum{2}) is  based on $f(Q)$ cosmologies but assuming the extra  friction term zero; the third one (SS\Romannum{3})  is based  on the $f(Q)$ model as well and using its true  extra friction term.
Then we could compare the SS\Romannum{1}$_\Lambda$ and SS\Romannum{2} to see the model effect, and compare SS\Romannum{2} and SS\Romannum{3} to see the effect of extra friction term.
The electromagnetic (EM)  data is  used as baseline in all models, including  direct determination of the Hubble parameters derived  from cosmic chronometer (CC) method \cite{Moresco:2020fbm,Favale:2023lnp},   baryon acoustic oscillations (BAO) of Dark Energy Spectroscopic Instrument (DESI) \cite{DESI:2024mwx} and the type Ia supernovae of PantheonPlus compilation (PantheonPlus) \cite{Scolnic:2021amr,Brout:2022vxf}.

The rest of this paper is organized as follows. In Section  \,\ref{ss1}, we will introduce the standard siren of gravitational wave.
In Section  \,\ref{data}, we will introduce the EM  observational data and the mock standard siren data.  In Section  \,\ref{model}, we will briefly describe the $f(Q)$ cosmologies.  In Section \,\ref{SS}, we will summarize the used data. In Section \,\ref{results}, we will report the constraint results. Finally, a brief summary will be given in Section \,\ref{summary}.


\section{The standard siren}\label{ss1}
The gravitational waves from compact systems are viewed as standard sirens to probe the evolution of the universe \cite{Holz:2005df,Schutz:1986gp}. From the GW signal, the luminosity distance $D_{L}^{SS}$ is measured directly, without invoking the cosmic distance ladder, since the standard sirens are self-calibrating. And it could be extracted from the GW amplitude
\begin{eqnarray}
	\label{dynamicalhA}
	h_A=\frac{4}{D_L^{SS}} (\frac{GM_c}{c^2})^{5/3}(\frac{\pi f_{GW}}{c})^{2/3},
\end{eqnarray}
where $ h_A$ is the SS amplitude, ``$A$" could be ``$+$" or ``$\times$", $D_L^{SS}$ is the luminosity distance for  gravitational wave standard siren, $M_c$ is the chirp mass, and $f_{GW}$ is the GW frequency. Obviously, the amplitude of gravitational wave standard siren will be affected by the background cosmological model.

Especially, the  propagation equation of standard siren in Fourier form is \cite{Cai:2016sby,Zhang:2021kqn}
\begin{eqnarray}
	\label{2}
	\overline{h}''_{A}+2\mathcal H [1+\delta (\eta)]\overline{h}'_{A} + k^2 \overline{h}_{A} = 0,
\end{eqnarray}
where $\overline{h}_{A}$ denotes the Fourier mode of the standard siren amplitude, the prime ``$\prime$" denotes a derivative with respect to conformal time $\eta$, and  $\mathcal H = a' /a$. And especially, $\delta$ is the so-called extra friction term which is zero  for the $\Lambda$CDM model or dark energy model in General Relativity.

In order to simplify the propagation equation within modified gravity theories,  by defining  a modified scale factor  $\tilde{a}'/{\tilde{a}} = \mathcal{H} [1+\delta(z)]$  and   $\chi_{A}=\tilde{a}h_{A}$,  we get \cite{Belgacem:2017ihm} 
\begin{eqnarray}
	\chi''_{A} + \left( k^{2} - \frac{\tilde{a}''}{\tilde{a}} \right) \chi_{A} = 0.
\end{eqnarray}
Then, the relation between the EM luminosity distance and the SS luminosity distance could be expressed as 
\begin{eqnarray}
	D^{SS}_{L}(z) = \exp \left( \int_{0}^{z} \frac{\delta(z')}{1+z'}dz' \right)D_{L}^{EM}(z),
	\label{delta}
\end{eqnarray}
where $D_{L}^{EM}(z)$ and $D^{SS}_{L}(z)$  are EM  and GW luminosity distances separately.
Obviously, the extra friction term $\delta$ characterizes the difference between the GW luminosity distance and the EM luminosity distance.
When $\delta$ is negative, $D^{EM}_{L}>D^{SS}_{L}$, there is a smaller $D^{SS}_{L}$ which denotes a larger Hubble parameter $H$, and then  a larger Hubble constant $H_0$ which  parameterizes the current expansion rate of our universe.

The $H_0$ parameter is related with the famous ``Hubble tension" problem. Cosmologically, $H_0$ could be measured from the cosmic microwave background which fit to a cosmological model such as $\Lambda$CDM (for instance, from Planck \cite{Planck:2018vyg}). And locally, $H_0$ could be measured from the observed redshift$-$distance relation in the Hubble flow for distant objects (for instance, from Cepheid variables and Type Ia supernovae by the SH0ES Team  based on the three-rung distance ladder method  \cite{Riess:2021jrx,Breuval:2024lsv, Murakami:2023xuy}). Explicitly, the Hubble tension refers to a  discrepancy of more than  $5\sigma$ between $H_0$ measured using these two measurements (see reviews by Refs. \cite{DiValentino:2021izs,Verde:2023lmm}).  
In cosmology, many  models which  could  return  to  $\Lambda$CDM  meet  Hubble tension problem as the $\Lambda$CDM does \cite{DiValentino:2020zio,Abdalla:2022yfr}.  Here, we will discuss Hubble tension for various $f(Q)$ cosmologies in the standard siren simulation.
The calculated values of Hubble tension could be denoted as \cite{Raveri:2018wln}
\begin{eqnarray}\label{R}
	T_{1}(\theta)=\frac{|\theta(D_{1})-\theta(D_{2})|}{\sqrt{\sigma_\theta^{2}(D_1)+\sigma_\theta^{2}(D_2)}}.
\end{eqnarray}
where $\theta$ is the best fitted values of $H_0$ from different data sets; the first data set $D_1$ represents the constraining results of cosmological fitting; the second data set $D_2$ is the chosen baseline measurement which is $H_{0} = 73.17 \pm 0.86 ~\mathrm{km/s/Mpc}$ from the latest SH0ES Team \cite{Riess:2021jrx,Breuval:2024lsv,Murakami:2023xuy}; and   $\sigma_\theta(D_1)$ and $\sigma_\theta(D_2)$ represent the  errors from $D_1$ and $D_2$ data sets respectively.

\section{The  method}\label{data}
The Markov Chain Monte Carlo (MCMC) package \texttt{CosmoMC} \cite{Lewis:2002ah} is employed to infer the  posterior probability distributions of parameters, and further to derive the best fitted values and their corresponding errors.  And  numerical simulation is also used to forecast results of surveys and targeted observations. We choose the Einstein Telescope as the representative of  third generation instruments which will detect thousands of Neutron Star Binary (NSB) and Black Hole Binary (BHB) mergers to probe the cosmic expansion at high redshifts      \cite{Zhao:2010sz,Cai:2016sby}. Here, we simulate the standard siren data by using the best fitted parameter values  from  EM combination and Einstein Telescope design index.

Firstly, we employ  current  EM observational data sets which are related to the cosmic distance, including the type Ia supernovae of PantheonPlus compilation (PantheonPlus) \cite{Scolnic:2021amr,Brout:2022vxf}, the direct measurements of the Hubble parameter derived  from cosmic chronometer method (CC) \cite{Moresco:2020fbm,Favale:2023lnp}, and baryon acoustic oscillations of Dark Energy Spectroscopic Instrument (DESI) \cite{DESI:2024mwx}, to perform the MCMC  analysis which give out the fiducial parameter values required by standard siren simulation.

The PantheonPlus compilation is acquired from $18$ distinct surveys \cite{Scolnic:2021amr,Brout:2022vxf}. It  contains $1701$ distance modulus data spanning within a redshift range of $0.00122<z<2.26137$.
The distance modulus, which is the observable quantity in the supernovae (SN) data,  is defined as
\begin{eqnarray}
	\mu=m-M=5\log_{10}[d_{L}(z)]+5\log_{10}[\frac{c/H_0}{Mpc}]+25,
\end{eqnarray}
where the luminosity distance is
\begin{eqnarray}
	\label{dEM}
	d_{L}=(1+z)\int^{z}_{0}\frac{d\tilde{z}}{H(\tilde{z})}.
\end{eqnarray}
In general, the goodness of fit for theoretical model is measured by  $\chi^2$ and likelihood functions ($L$) which is  expressed as $\chi^2  = -2{\rm ln} L $. To get best fit, the value of $\chi^2$ is needed to minimize.
In the context of the PantheonPlus compilation, the $\chi^2$ likelihood
function could be computed by
\begin{eqnarray}
	\chi^{2}_{PantheonPlus}=\sum_{i=1}^{1701}\Delta \mu^T  \mathbf{C}^{-1}_{stat+sys}  \Delta \mu,
\end{eqnarray}
where the covariance matrix ($\mathbf{C}_{stat+sys}$)  includes both the systematic and statistical errors,  $\Delta \mu$ is the vector of $1701$ SN distance modulus residuals computed as
\begin{equation}
	\Delta \mu_i = \mu^{model}(z_i)-\mu^{data}_i.
\end{equation}
 
The Hubble parameter data could be  obtained through the cosmic chronometer (CC) method
which calculate the differential ages of passively evolving galaxies. Here, the used CC compilation contains $32$ data points  \cite{Zhang:2012mp,Jimenez:2003iv,Moresco:2012by,Moresco:2016mzx,Stern:2009ep,Moresco:2015cya,Simon:2004tf,Chuang:2011fy,Gomez-Valent:2018hwc,Borghi:2021rft} which are tabulated in Table 1 of Ref.\cite{Favale:2023lnp}. And 
we use the covariance matrix for computations as described in Ref. \cite{Moresco:2020fbm}. Then, 
the form of $\chi^2$ of Hubble parameter data through the cosmic chronometer (CC) method  is
\begin{eqnarray}
	\label{chi2}
	\chi^2_{CC}=\sum_{i=1}^{32}\Delta H(z)^T  \mathbf{C}^{-1}_{stat+model+young+met}  \Delta H(z),
\end{eqnarray}
where  $\Delta H(z)=H(z_i)^{model}-H^{data}_{i}$, ``stat'', ``young'', ``model'' and ``met'' denote the contributions to the covariance due to statistical errors, young component contamination, dependence on the chosen model, and stellar metallicity respectively.

Furthermore, the properties of BAO are derived from the matter power spectrum which are related with the matter fluctuation perturbation. In the clustering of matter of late universe, they could serve as a standard ruler to map the expansion history of the universe.  Here, we adopt the first-year data released by the DESI collaboration \cite{DESI:2024mwx}, which includes observations from four different classes of extragalactic targets: the bright galaxy sample (BGS) \cite{Hahn:2022dnf}, luminous red galaxies (LRG) \cite{DESI:2022gle}, emission line galaxies (ELG) \cite{Raichoor:2022jab}, and quasars (QSO)  \cite{Chaussidon:2022pqg}.
The DESI provides robust measurements of the transverse comoving distance ($D_M$),the Hubble distance ($D_H$) and the angle-average distance ($D_V$) relative to the drag-epoch sound horizon ($r_d$) in seven redshift bins from over 6 million extragalactic objects.
The DESI data are summarized in Table 1 of Ref. \cite{DESI:2024mwx}. Firstly, we calculate the $\chi^2$ related to the BGS  and  QSO data as below 
\begin{eqnarray}
	\label{chi2}
	\chi^2_{DESI1}=\sum_{i=1}^{2}\frac{((D_V / r_d)^{model}-(D_V / r_d)^{data})^{2}}{\sigma^2_{z_i}}.
\end{eqnarray}
And, the data of $D_M / r_d$ and $D_H / r_d$ from tracers LRG1, LRG2, LRG3+ELG1, ELG2, and Lya QSO,  degenerate at the same redshift $z$. 
Following Ref.\cite{strang2000introduction,Li:2024hrv}, the data vector $D$ could be constructed as
\begin{equation}
	\label{D}
	D \equiv \begin{pmatrix}
		D_M / r_d \\
		D_H / r_d \\
	\end{pmatrix},
\end{equation}
with its covariance matrix defined as \cite{strang2000introduction,Li:2024hrv}:
\begin{equation}
	\text{Cov}_{DESI2} = \begin{bmatrix}
		\sigma_1^2 & r \cdot \sigma_1 \cdot \sigma_2 \\
		r \cdot \sigma_1 \cdot \sigma_2 & \sigma_2^2
	\end{bmatrix},
\end{equation}
where $\sigma_1$ and $\sigma_2$ denote the standard deviations of $D_M / r_d$ and $D_H / r_d$ respectively. The correlation coefficient between $D_M / r_d$ and $D_H / r_d$, which is denoted as $r$, is provided in Table 1 of Ref. \cite{DESI:2024mwx}.
Then the second  $\chi^2_{DESI2}$ is expressed as
\begin{eqnarray}
	\chi^2_{DESI2} = \sum_i^{10} \Delta \text{D}_i^T  \text{Cov}_{DESI2}^{-1} \Delta \text{D}_i,
\end{eqnarray}
where $\Delta \text{D}_i = D_i^{model} - D_i^{data}$ is the data vector constructed by Eq. (\ref{D}).

We combine   the PantheonPlus, CC and DESI observational data as EM compilation which has $1745$ data points. And the best fitted values of EM compilation, which correspond to the minimum of sum of $\chi^{2}_{PantheonPlus}$, $\chi^{2}_{CC}$ and $\chi^{2}_{DESI}$, is used as fiducial parameter values in standard siren simulations.

Meanwhile, the strain $h(t)$ in the gravitational wave interferometers could be written as \cite{Zhao:2010sz}
\begin{eqnarray}
	h(t)=F_{+} (\theta, \phi, \psi )h_{+}(t) + F_{\times}(\theta, \phi, \psi)h_{\times}(t),
\end{eqnarray}
where $F_+$, $F_{\times}$ are the antenna pattern functions sensed by the gravitational wave detector. The redshift range is chosen as $0<z<5$.
And  the standard siren sources considered in this work include the merger events from black hole-neutron star  systems and binary neutron star systems, both of which are expected to exhibit afterglows in the EM radiation after they emit  bursts of gravitational wave. Thus, BNS and BHNS could be observed  not only  as a transient standard siren event, but also as an EM counterpart, and could be used as standard siren candidates.

In this work, to calculate the errors of the simulated data, we utilize the  one-sided noise power spectral density (PSD) which characterizes performance of the gravitational wave detector. The measurement errors of luminosity distance is also related to the weak lensing effects. Following the studies in Refs.\,\cite{ Zhao:2010sz,Li:2013lza},  this weak lensing error is assumed to be $0.05z$. Thus, the total uncertainty on the measurement of $D_L$ is taken as
\begin{eqnarray}
	\sigma_{D_L}=\sqrt{\sigma^{2}_{\rm inst}+\sigma^{2}_{\rm lens}}=\sqrt{(\frac{2D_{L}}{\rho})^{2}+(0.05zD_{L})^{2}},
\end{eqnarray}
where $\sigma^{2}_{\rm inst}$ is the instrumental error calculated by Fisher Matrix   and  $\rho$ is the ratio of signal to noise which is usually chosen as $\rho >8$.  
In this paper, we will simulate $1000$  standard siren data points expected to be detected by Einstein Telescope in its 10-year observation. To achieve this, the Fisher matrix approach is utilized \cite{Cai:2016sby}.
And, we roughly assume that there are 500 BNS events and 500 BHNS events. 
The $\chi^2$ of SS data could be expressed as below:
\begin{eqnarray}
	\chi^2_{SS}=\sum_{i=1}^{1000}\frac{(D_L^{model}(z_i)-(D_L^{data} )_i)^{2}}{\sigma^2_{z_i}}.
\end{eqnarray}
Then, the standard sirens method offers a new independent way to probe the cosmic expansion.

\section{The $f(Q)$ cosmologies}\label{model}
In the Symmetric Teleparallel gravity, the  non-metricity $Q$, which represents the variation in length of a vector during parallel transport,  is  used to describe the gravitational interaction.   And, 
its natural extension, the  $f(Q)$ modified gravity,  has revealed many interesting cosmological phenomena  as shown in the literature.

The action of $f(Q)$ cosmology could be given by \cite{BeltranJimenez:2017tkd,BeltranJimenez:2019tme,Heisenberg:2023lru},
\begin{eqnarray}
	\label{action00}
S=\int\sqrt{-g} \left[ -\frac{1}{16\pi G}f(Q)+\mathcal{L}_m\right]~d^4x,
\end{eqnarray}
where $f(Q)$ is an arbitrary function of the  non-metricity scalar $Q$;  $\mathcal{L}_m$ is the matter Lagrangian density and $g$ is the
determinant of metric $g_{\mu\nu}$.

As the homogeneous and isotropic Friedmann-Robertson-Walker (FRW) spacetime  is considered,
  we obtain the exact value of non-metricity scalar 
\begin{eqnarray}
	Q=6H^{2}.
\end{eqnarray}
Here, $H={\dot{a}}/{a}$ is the Hubble parameter.
Note that in the $f(Q)$ gravity, the non-metricity scalar $Q$  plays the role of Ricci scalar $R$ in GR, which indicates  these two categories of modified gravity theories  ($f(Q)$ and $f(R)$) are equivalent in the background level. It is convenient to set $f(Q)=Q+F(Q)$ where the $F(Q)$ part  represents the cosmic acceleration effect.
Then the Friedmann equations for flat spacetime take the following form \cite{BeltranJimenez:2017tkd, BeltranJimenez:2019tme,Heisenberg:2023lru}
\begin{eqnarray}
&&3H^{2}=\rho+\frac{F}{2}-QF_Q \,, \label{eq:FRDEQ1}\\
&&\dot{H}=\frac{F-Q-2QF_Q}{4(2Q F_{QQ}+F_Q+1)} \,,\label{eq:FRDEQ2}
\end{eqnarray}
where $\rho$ and $p$  are the energy density and pressure for the matter fluid, and satisfy conservation equation:
\begin{eqnarray}
		\dot{\rho}+3H(1+w)\rho =0\,,\label{eq:cons}
\end{eqnarray}
 where $w$ is the equation-of-state (EoS) parameter.

Correspondingly, the effective energy density $\rho_{\mathrm{eff}}$ and effective pressure $p_{\mathrm{eff}}$ for the acceleration part that is sourced from $F(Q)$ could be described as
\begin{eqnarray}
		\rho _{\mathrm{eff}} &=& \frac{F}{2}-QF_{Q}\,, \\
		p_{\mathrm{eff}} &=& 2\dot{H}(2QF_{QQ}+F_Q)-\rho_{\mathrm{eff}}
\,.
\end{eqnarray}
Thus, the effective EoS parameter could be given by
	\begin{eqnarray}
	w_{\mathrm{eff}}=-1+\frac{1-1/(F/Q-2F_Q)}{1+1/(2QF_{QQ}+F_Q)}.
\end{eqnarray}

In the framework of $f(Q)$ gravity, the extra friction term in Eq. (\ref{2}) takes the below form: 
 	\begin{eqnarray}
\delta (z)=\frac{d\ln f_Q}{2\mathcal H d\eta}.
\end{eqnarray}
And the dimensionless Hubble parameter  could be  expressed   as
\begin{eqnarray}
E(z) =  \frac{H}{H_{0}},
\end{eqnarray}
where the subscript ``0" denotes the present time. 

Especially, the  $\Lambda$CDM model could be regarded  as $F(Q)=\Lambda$ where $w_\mathrm{{eff}}=-1$ and $\delta=0$. 
Here, for convenience, the constraining models are divided  into the $\Lambda$CDM-like one (the power-law $f(Q)_{P}$   \cite{Bengochea:2008gz,Chen:2010va} and  square-root exponential $f(Q)_{E}$ \cite{Linder:2010py} models) and non  $\Lambda$CDM-like one (the power exponential $f(Q)_{PE}$    \cite{Anagnostopoulos:2021ydo,Ferreira:2023awf} and  hyperbolic tangent $f(Q)_{HT}$  \cite{Wu:2010av,Anagnostopoulos:2022gej,Qi:2017xzl} models). Next, we will introduce them one by one.

\subsection{The power-law form: $f(Q)_{P}$ model}
The power-law $f(Q)$ model \cite{Bengochea:2008gz,Chen:2010va} (hereafter $f(Q)_{P}$ model)   is a simple and notable model,
\begin{eqnarray}
	F(Q)=\alpha Q^{b},
\end{eqnarray}
where $ \alpha=Q_{0}^{1-b}(1-\Omega_{m0})/(1-2b)$, and $b$ is the new freedom which quantifies deviation from  $\Lambda$CDM model.
When $b=0$, this model degenerates to $\Lambda$CDM model. When $b=1/2$, this model reduces to the Dvali-Gabadadze-Porrati (DGP) model \cite{Dvali:2000hr}. When $b=1$ the model is the same as the standard cold dark matter model after re-scaling the Newton’s constant. Then it is required that $b<1$ in order to obtain an accelerating cosmic expansion.
Ref. \cite{Zhang:2021kqn} gave out that the $f(Q)_{P}$ model could be distinguished from $\Lambda$CDM model by using GW and EM combined data. This phenomenon is worthy checking by the updating data.

\subsection{The square-root exponential form: $f(Q)_{E}$ model}
And, we introduce the square-root exponential model   \cite{Linder:2010py}  (hereafter f(Q)$_{E}$ model )
\begin{eqnarray}
	F(Q)=\alpha Q_0(1-e^{-p\sqrt{Q/Q_0}}),
\end{eqnarray}
where $\alpha=(1-\Omega_{m0})(1-(1+p)e^{-p}) $ and   $p$  is  model parameter.
Usually,  we set $b=1/p$ for convenience. As the $f(Q)_P$  model, the $f(Q)_E$ model comes back to $\Lambda$CDM model  when $b=0$.  And  $b\neq 0$ in $f(Q)_E$ model indicates an essential deviation from  $\Lambda$CDM model. However, $b\to+0$  corresponds to $p\to +\infty$, while $b\to-0$  corresponds to $p\to-\infty$.
Then, getting  across $b=0$ means crossing the singularity $p$. And when $b<0$, $e^{-p\sqrt{Q/Q_0}}$  grows exponentially. Therefore, to avoid the singularity, we set  the prior $b>0$ for all constraints which is favored by Ref \cite{Basilakos:2018arq}.
In the literature \cite{Zhang:2021kqn}, the EM constraint with Pantheon included shows the  Hubble tensions as large as $3.21\sigma$. So we constrain it with PantheonPlus included data to see whether the Hubble tension could be alleviated.

\subsection{The power exponential form: $f(Q)_{PE}$ model}\label{sec:exp_model}
In Refs.\,\cite{Anagnostopoulos:2021ydo,Ferreira:2023awf}, a power exponential form of $f(Q)$ model has been introduced (hereafter $f(Q)_{PE}$  model), and it could be expressed as:
\begin{eqnarray}
	F(Q)= Q \left(e^{ \lambda \frac{Q_0}{Q} }-1 \right),
\end{eqnarray}
where the derived parameter $\lambda$ which is determined by $\Omega_{m0}$  could be described as
\begin{eqnarray}
	\lambda =\frac{1}{2} + W_{0}\left( -\frac{\Omega_{m0}}{2e^{\frac{1}{2}}} \right),
\end{eqnarray}
and $W_{0}$ is the Lambert function.
Similar to the $\Lambda$CDM  model, the $f(Q)_{PE}$  model has two free parameters $\Omega_{m0}$ and $H_0$.
Note  that, the $f(Q)_{PE}$ model theoretically could not come back to the $\Lambda$CDM for any values of $\lambda$, thus it is  called the non $\Lambda$CDM-like model  here.

At high-redshift where $Q_0 \ll Q$,  $F_Q\simeq-\lambda^2Q_0^2/Q^2 $ and $F_{QQ} \simeq\lambda^2Q_0^2/Q^3$.
The effective EoS can be expanded as
\begin{eqnarray}
	\label{w1}
	w_{\mathrm{eff}}=-1-\lambda\frac{Q_0}{Q}.
\end{eqnarray}
If $\lambda>0$, the $w_{\mathrm{eff}}$ would  approach to $-1$ from the phantom side. And the  extra friction term could be expressed as 
\begin{eqnarray}
	\label{deltae}
	\delta = -\frac{3}{2}\lambda^{2} \left( \frac{Q_{0}}{Q} \right)^{2},
\end{eqnarray}
which tends to $0$ from negative side. 

The power exponential model has the same free parameters with $\Lambda$CDM model, but does not degenerate with $\Lambda$CDM model. As it has been proved to  alleviate  the Hubble tension \cite{Anagnostopoulos:2021ydo,Ferreira:2023awf}, it will be interesting to constrain this model.

\subsection{The hyperbolic tangent form: $f(Q)_{HT}$ model}
For the purpose of realizing the crossing of the phantom divide line \cite{Wu:2010av},   the  hyperbolic tangent form of the $f(Q)$ model is proposed as \cite{Anagnostopoulos:2022gej,Qi:2017xzl} (hereafter $f(Q)_{HT}$ model):
\begin{eqnarray}
	\label{ModelI}
	F(Q)=\alpha Q_{0} \left( \frac{Q_0}{Q}\right)^{-b}\tanh \frac{Q_0}{Q},
\end{eqnarray}
where $b$ is an additional free parameter compared with $\Lambda$CDM model or $f(Q)_{PE}$ model. And the dimensionless parameter $\alpha$ could be expressed as
\begin{eqnarray}
	\alpha = \frac{ 1-\Omega_{m0}}{(1-2b) \tanh{1} + 2\sech^{2}{1}}.
\end{eqnarray}
The $f(Q)_{HT}$ model could  not come back to the $\Lambda$CDM for any values of $b$ as well. 

When $Q\gg Q_0$, $\tanh (Q_0/Q) \simeq Q_0/Q$ and $\sech (Q_0/Q) \simeq 1$. Then,  $F_{Q}\simeq \alpha(b-1)(Q_0/Q)^{(2-b)}$ and $F_{QQ}\simeq \alpha(1-b)(2-b)(Q_0/Q)^{(2-b)}/Q$ which are small as well at  high redshift. And, the EoS and extra friction are approximately to be
\begin{eqnarray}
	&&	w_{\mathrm{eff}} = -2+b, \label{w2} \\
	&&	\delta=- \alpha(1-b)(2-b) \left(\frac{Q_{0}}{Q} \right)^{2-b}.
	\label{deltah}
\end{eqnarray}

As the effective energy density must be larger than $0$,  we need to give a prior of $b$ for  the model. So, we  divide the Hyperbolic Tangent model to  two branches:
\begin{description}
	\item[\textbf{$f(Q)_{HT1}$:}] We set  a prior $b<0.5$ for this model. The $w_{\mathrm{eff}}$ would  approach to the phantom side at high redshifts. The $\delta$   tends to $0$ from negative side.
	\item[\textbf{$f(Q)_{HT2}$:} ]As Big Bang Nucleosynthesis (BBN) give the constraint $b<1.946$ \cite{Kavya:2024ssu} \footnote{Another BBN constraints for the $f(Q)_{HT2}$ model are satisfied  $b \lesssim1.88$    \cite{Anagnostopoulos:2022gej}.}, we set  a prior $1.500<b<1.946$ for this model. At high redshifts, the $w_{\mathrm{eff}}$ would  approach to the quintessence side, while the $\delta$   tends to $0$ from positive side. 
\end{description}

Meanwhile, under the  constraining of  existing data,  the hyperbolic tangent ($f(Q)_{HT}$) model   is ``punished"  by Akaike information criterion (AIC) and Bayesian information criterion  (BIC)  \cite{Xu:2018npu,Briffa:2021nxg} which worth further study.

\section{The simulted standard siren data}\label{SS}
 And,  we summary the used data as below:
\begin{description}
	\item[\textbf{EM}:] The observational data of  PantheonPlus, CC and DESI  are combined as EM compilation which has $1745$ data. Its best fitted values is used as baseline parameter values in standard siren simulations. And we list the constraining results of EM in Table \ref{tab}.
	
	\item[\textbf{SS\Romannum{1}$_{\Lambda}$: }]  The SS\Romannum{1}$_{\Lambda}$ simulation used in $\Lambda$CDM  and $f(Q)$ models is based on the $\Lambda$CDM model with $\Omega_{m0}=0.322$ and  $H_0=73.23$. The ``SS" is the abbreviation of standard siren. And the subscript ``$\Lambda$" denotes the $\Lambda$CDM fiducial model. This simulation is mainly used for testing model effect in $f(Q)$ models.
	
	\item[\textbf{SS\Romannum{2}: }]  As Table \ref{tab}  shows, we use $f(Q)_{P}$ model with $\Omega_{m0}=0.327$, $H_0=73.35$ and $b=-0.082$ for the SS\Romannum{2}$_{P0}$ simulation; use $f(Q)_{E}$ model with $\Omega_{m0}=0.322$, $H_0=73.19$ and $b=0.106$ for the SS\Romannum{2}$_{E0}$ simulation; use
	$f(Q)_{PE}$ model with $\Omega_{m0}=0.347$ and $H_0=73.78$ for the SS\Romannum{2}$_{PE0}$ simulation; 
	use $f(Q)_{HT1}$ model with $\Omega_{m0}=0.336$, $H_0=72.65$ and $b=0.218$ for the SS\Romannum{2}$_{HT10}$ simulation;  and use $f(Q)_{HT2}$ model with $\Omega_{m0}=0.322$, $H_0=73.37$ and  $b=1.624$  for  the SS\Romannum{2}$_{HT20}$ simulation. Especially, $\delta=0$ is  assumed for all the SS\Romannum{2} simulations. Physically, the SS\Romannum{2} simulations do not correspond to any true data. This simulation is used to denote the model and extra friction effects by comparing with SS\Romannum{1}$_{\Lambda}$ and SS\Romannum{3}. The subscripts ``P", ``E", ``PE", ``HT1" and ``HT2" denotes the simulated model. And the subscript ``$0$" denotes $\delta=0$.
	
	\item[\textbf{SS\Romannum{3}: }] The model parameter values are the same as SS\Romannum{2}  except that  we use  $\delta_0=-0.024$,  $\delta_0=0.005$,  $\delta_0=-0.092$, $\delta_0=0.048$ and $\delta_0=-0.415$ which are the EM constraining results listed in Table \ref{tab} for  the  SS\Romannum{3}$_{P}$, SS\Romannum{3}$_{E}$,  SS\Romannum{3}$_{PE}$, SS\Romannum{3}$_{HT1}$ and SS\Romannum{3}$_{HT2}$   simulations separately.    
\end{description}

Here, the EM and SS\Romannum{1}$_{\Lambda}$  data will be applied to all the models. While as SS\Romannum{2} and SS\Romannum{3} simulations are based on different $f(Q)$ model, they will be applied to their fiducial model. To compare the real and simulated data we will plot the evolution of luminosity distances. And, we will use the dashed/dotted lines for SS\Romannum{1}$_{\Lambda}$/SS\Romannum{2} in $f(Q)$ cosmologies. While we use the solid lines for all the EM  data, for the simulated data SS\Romannum{1}$_{\Lambda}$ in $\Lambda$CDM model and for the simulated data SS\Romannum{3} in $f(Q)$ models. 

And to determine the most suitable model according to data, we will  make a comparison for  the  EM and SS\Romannum{1}$_{\Lambda}$ results by using  the minimum of $\chi^2$ value which  stands for the best fit. 
However, a higher number of parameters can artificially improve the fit, leading to a smaller $\chi^2$, making it unreliable for model comparison. To address this issue, we employ the Akaike information criterion (AIC) where $ AIC =  \chi^2 + 2N$ with $N$ as the number of free parameter \cite{Akaike:1974vps} and the Bayesian information criterion (BIC) where $BIC = \chi^2 + N \ln{m}$ with $m$ as the number of data points used in the fit \cite{Schwarz:1978}. Here,  the $\chi^2_{SS}$s should be around the number of  data $1000$. And, for Gaussian errors, the difference between two models could be written as  $\Delta AIC = \Delta \chi^2 + 2\Delta N$. 
Similar to the AIC, the difference denoted by BIC has the form 
$\Delta  BIC = \Delta  \chi^2+ \Delta  N \ln{m}$. The $\Delta AIC=5$($\Delta BIC\geq2$ )and $\Delta AIC=10$($\Delta BIC\geq6$) are considered to be the positive and strong evidence against the weaker model.

\begin{table*}
	\small
	\begin{center}
		\caption{The best fitted values  with 1$\sigma$ and 2$\sigma$ standard errors from the constraints of EM and  SS related data for the $\Lambda$CDM and $f(Q)$ models. The best fitted values of EM is used as baseline parameter values in standard siren simulations. Explicitly, the best fitted EM values of $\Lambda$CDM model are used for SS\Romannum{1}$_{\Lambda}$ simulation. The best fitted EM values of $f(Q)$ models are used for SS\Romannum{2} and SS\Romannum{3} simulations. Especially, we list the Hubble tension at the end of the table where the reference Hubble constant value $H_{0} = 73.17 \pm 0.86 ~\mathrm{km/s/Mpc}$ is from the SH0ES Team \cite{Riess:2021jrx,Breuval:2024lsv} as Section \ref{ss1} stated. We did not list the Hubble tensions of SS\Romannum{1}$_{\Lambda}$ and SS\Romannum{2} data for $f(Q)$ models because they are not physical.}	
		\renewcommand\arraystretch{1.5}
		\begin{tabular}{|c|c|cccccc|}
			\hline
			Model& Data&$\Omega_{m0}$ & $H_0 (\rm km/s/Mpc)$ & $b$& $\delta_0$& $w_{\rm{eff0}}$ &  Hubble tension  \\
			\hline
			\multirow{2}*{$\Lambda$CDM} & EM& $0.322^{+0.011+0.023}_{-0.011-0.022}$  & $73.23^{+0.17+0.34}_{-0.17-0.34}$ & $-$ & $0$ &  $-1$ & $0.07\sigma$ \\
			&SS\Romannum{1}$_\Lambda$ & $0.316^{+0.014+0.028}_{-0.014-0.027}$ & $73.56^{+0.54+1.11}_{-0.54-1.11}$ & $-$ & $0$ & $-1$ & $0.38\sigma$  \\	
			
			\hline
			\multirow{4}*{$f(Q)_{P}$}
			&EM& $0.327^{+0.012+0.024}_{-0.012-0.023}$  & $73.35^{+0.19+0.38}_{-0.19-0.38}$ & $-0.082^{+0.061+0.111}_{-0.053-0.122}$& $-0.024^{+0.017+0.033}_{-0.017-0.032}$ & $-1.025^{+0.017+0.033}_{-0.017-0.035}$ & $0.20\sigma$\\
			&SS\Romannum{1}$_\Lambda$ &$0.281^{+0.059+0.084}_{-0.031-0.101}$  & $73.09^{+0.88+1.81}_{-0.88-1.70}$ & $0.201^{+0.401+0.601}_{-0.252-0.672}$ & $0.153^{+0.073+0.481}_{-0.252-0.332}$  & $-0.915^{+0.121+0.241}_{-0.121-0.230}$ & $-$\\
			&SS\Romannum{2}$_{P0}$  &$0.330^{+0.036+0.054}_{-0.019-0.066}$ 
			&$73.74^{+0.91+1.61}_{-0.80-1.71}$ 
			&$-0.211^{+0.421+0.642}_{-0.350-0.711}$ &$-0.032^{+0.036+0.221}_{-0.131-0.150}$ &$-1.048^{+0.072+0.202}_{-0.130-0.181}$ & $-$\\
			&  SS\Romannum{3}$_{P}$ &$0.383^{+0.028+0.047}_{-0.018-0.053}$  & $73.88^{+0.86+1.51}_{-0.73-1.59}$ & $-0.420^{+0.281+0.599}_{-0.442-0.580}$& $-0.090^{+0.022+0.160}_{-0.096-0.101}$ & $-1.118^{+0.049+0.190}_{-0.131-0.151}$ & $0.61\sigma$\\
			
			\hline
			\multirow{4}*{$f(Q)_{E}$}
			&EM& $0.322^{+0.011+0.023}_{-0.011-0.022}$& $73.19^{+0.18+0.35}_{-0.18-0.37}$ & $0.106^{+0.035+0.103}_{-0.106-0.106}$& $0.005^{+0.001+0.018}_{-0.004-0.005}$ & $-0.992^{+0.001+0.030}_{-0.007-0.008}$ & $0.02\sigma$\\
			&SS\Romannum{1}$_\Lambda$ &$0.289^{+0.038+0.054}_{-0.017-0.070}$  
			&$72.55^{+1.11+1.70}_{-0.88-1.91}$ 
			&$0.424^{+0.077+0.577}_{-0.424-0.424}$ 
			&$0.098^{+0.048+0.171}_{-0.097-0.098}$  
			&$-0.861^{+0.073+0.221}_{-0.139-0.139}$ & $-$\\
			&SS\Romannum{2}$_{E0}$  &$0.306^{+0.048+0.066}_{-0.025-0.084}$ 
			&$71.23^{+1.00+2.01}_{-1.00-1.91}$ 
			&$0.572^{+0.209+0.726}_{-0.438-0.572}$ &$0.146^{+0.078+0.181}_{-0.143-0.146}$ &$-0.791^{+0.141+0.231}_{-0.161-0.209}$ & $-$\\
			&  SS\Romannum{3}$_{E}$ &$0.327^{+0.019+0.036}_{-0.015-0.038}$  & $72.36^{+0.78+1.31}_{-0.60-1.40}$ & $0.220^{+0.056+0.277}_{-0.220-0.220}$& $0.038^{+0.013+0.101}_{-0.036-0.038}$ & $-0.941^{+0.024+0.151}_{-0.059-0.059}$ & $0.74\sigma$\\
			\hline
			\multirow{4}*{$f(Q)_{PE}$} & EM &  $0.347^{+0.012+0.024}_{-0.012-0.023}$  & $73.78^{+0.18+0.35}_{-0.18-0.36}$ & $-$ & $-0.092^{+0.001+0.002}_{-0.001-0.002}$&  $-1.132^{+0.004+0.008}_{-0.004-0.008}$& $0.69\sigma$ \\
			& SS\Romannum{1}$_\Lambda$ &$0.336^{+0.013+0.027}_{-0.013-0.025}$ & $74.52^{+0.54+1.11}_{-0.54-1.11}$ & $-$ & $-0.091^{+0.001+0.002}_{-0.001-0.002}$& $-1.129^{+0.005+0.009}_{-0.005-0.009}$ & $-$ \\
			&SS\Romannum{2}$_{PE0}$ &$0.339^{+0.016+0.032}_{-0.016-0.029}$  & $74.37^{+0.67+1.31}_{+0.67-1.31}$ & $-$ & $-0.091^{+0.001+0.003}_{-0.001-0.002}$ & $-1.130^{+0.005+0.010}_{-0.005-0.011}$ & $-$ \\	
			&  SS\Romannum{3}$_{PE}$ &$0.404^{+0.017+0.033}_{-0.017-0.031}$  & $73.94^{+0.62+1.21}_{-0.62-1.21}$ & $-$ & $-0.094^{+0.001+0.001}_{-0.001-0.001}$ & $-1.151^{+0.005+0.010}_{+0.005-0.010}$ & $0.73\sigma$ \\
			
			\hline
			\multirow{4}*{$f(Q)_{HT1}$}
			&EM& $0.336^{+0.012+0.023}_{-0.012-0.022}$  & $72.65^{+0.24+0.46}_{-0.24-0.47}$ & $0.218^{+0.045+0.081}_{-0.038-0.084}$& $0.048^{+0.024+0.056}_{-0.028-0.050}$ & $-0.827^{+0.042+0.099}_{-0.050-0.086}$ & $0.58\sigma$\\
			&SS\Romannum{1}$_\Lambda$ &$0.349^{+0.016+0.033}_{-0.016-0.031}$  & $72.30^{+1.50+2.41}_{-1.21-2.70}$ & $0.171^{+0.291+0.329}_{-0.097-0.431}$ & $0.070^{+0.081+0.311}_{-0.171-0.242}$  & $-0.777^{+0.131+0.572}_{-0.310-0.421}$ & $-$\\
			&SS\Romannum{2}$_{HT10}$  &$0.345^{+0.018+0.035}_{-0.018-0.034}$ 
			&$71.07^{+1.21+2.71}_{-1.61-2.52}$ 
			&$0.319^{+0.181+0.181}_{-0.030-0.318}$ &$0.190^{+0.131+0.302}_{-0.201-0.272}$ &$-0.558^{+0.231+0.571}_{-0.392-0.491}$ &$-$\\
			&  SS\Romannum{3}$_{HT1}$ &$0.435^{+0.019+0.037}_{-0.019-0.036}$  & $70.96^{+0.96+2.20}_{-1.31-2.01}$ & $0.310^{+0.190+0.190}_{-0.035-0.311}$& $0.164^{+0.141+0.261}_{-0.161-0.242}$ & $-0.542^{+0.281+0.562}_{-0.362-0.492}$ & $1.56\sigma$\\
			\hline
			\multirow{4}*{$f(Q)_{HT2}$}
			&EM &$0.322^{+0.012+0.025}_{-0.012-0.024}$  & $73.37^{+0.18+0.35}_{-0.18-0.36}$ & $1.624^{+0.016+0.032}_{-0.016-0.032}$ & $-0.415^{+0.023+0.041}_{-0.021-0.045}$ & $-0.988^{+0.015+0.030}_{-0.015-0.030}$ & $0.23\sigma$\\
			&SS\Romannum{1}$_\Lambda$& $0.265^{+0.044+0.084}_{-0.054-0.081}$  & $73.06^{+0.63+1.41}_{-0.71-1.30}$ & $1.778^{+0.161+0.168}_{-0.051-0.193}$& $-0.238^{+0.170+0.201}_{-0.078-0.231}$ & $-0.839^{+0.140+0.171}_{-0.073-0.202}$ & $-$\\
			&SS\Romannum{2}$_{HT20}$  &$0.327^{+0.048+0.064}_{-0.021-0.085}$ & $73.86^{+0.49+0.94}_{-0.49-0.99}$ & $1.629^{+0.029+0.191}_{-0.129-0.129}$ &$-0.415^{+0.069+0.231}_{-0.161-0.190}$ &$-0.985^{+0.038+0.190}_{-0.121-0.132}$ & $-$\\
			&SS\Romannum{3}$_{HT2}$  &$0.398^{+0.002+0.002}_{-0.001-0.004}$ 
			&$84.74^{+0.33+0.65}_{-0.33-0.63}$ 
			&$1.944^{+0.002+0.002}_{-0.001-0.004}$ &$-0.031^{+0.002+0.003}_{-0.001-0.005}$ &$-0.483^{+0.004+0.005}_{-0.001-0.009}$ & $12.56\sigma$\\
			\hline
			
		\end{tabular}		
		\label{tab}		
	\end{center}	
\end{table*}

\begin{table*}
	\small
	\begin{center}
		\caption{The $\chi^2$ , AIC and BIC values of the PantheonPlus, CC,  DESI, EM and SS\Romannum{1}$_\Lambda$ data for $\Lambda$CDM model and $f(Q)$ cosmologies. Because SS\Romannum{2} and SS\Romannum{3} are simulated based on different $f(Q)$ models, we could not compare their $\chi^2$s (AICs, BICs). Then, the $\chi^2$s (AICs, BICs) of SS\Romannum{2} and SS\Romannum{3} are not list here. }	
		\renewcommand\arraystretch{1.5}
		\begin{tabular}{|c|ccccccccc|}
			\hline
			\multirow{7}*{EM}
			&Model&$\chi^{2}_{PantheonPlus}$&$\chi^{2}_{CC}$&$\chi^{2}_{DESI}$&$\chi^2_{EM}$&$AIC_{EM}$&$BIC_{EM}$&$\Delta AIC_{EM}$&$\Delta BIC_{EM}$\\
			&$\Lambda$CDM&$1758.9$&$17.6$&$22.6$&$1799.1$&$1803.1$&$1814.0$&$0$&$0$\\
			
			&$f(Q)_{P}$&$1761.6$&$17.5$&$20.3$&$1799.4$  &$1805.4$&$1821.8$&$2.3$&$7.8$\\
			
			&$f(Q)_{E}$&$1758.3$&$17.6$&$24.6$&$1800.5$  &$1806.5$&$1822.9$&$3.4$&$8.9$\\
			
			&$f(Q)_{PE}$&$1774.6$&$17.9$&$29.2$&$1821.7$ &$1825.7$&$1836.6$&$22.6$&$22.6$ \\
			
			&$f(Q)_{HT1}$&$1750.4$&$17.3$&$14.6$&$1782.3$&$1788.3$&$1804.7$&$-14.8$&$-9.3$\\
			
			&$f(Q)_{HT2}$&$1763.7$&$17.2$&$22.5$&$1803.4$&$1809.4$&$1825.8$&$6.3$&$11.8$   \\	
			\hline
			\multirow{7}*{$SS\Romannum{1}_\Lambda$}
			&Model&$-$&$-$&$-$&$\chi^{2}_{SS\Romannum{1}_\Lambda}$&$ AIC_{SS\Romannum{1}_\Lambda}$&$BIC_{SS\Romannum{1}_\Lambda}$&$\Delta AIC_{SS\Romannum{1}_\Lambda}$&$\Delta BIC_{SS\Romannum{1}_\Lambda}$\\
			
			&$\Lambda$CDM	&$-$&$-$&$-$&$983.9$&$987.9$&$997.7$&$0$&$0$ \\
			
			&$f(Q)_{P}$	&$-$&$-$&$-$&$984.1$&$990.1$&$1004.8$&$2.2$&$7.1$\\
			
			&$f(Q)_{E}$	&$-$&$-$&$-$&$984.3$&$990.3$&$1005.0$&$2.4$&$7.3$\\
			
			&$f(Q)_{PE}$	&$-$&$-$&$-$&$986.9$&$990.9$&$1000.7$&$3.0$&$3.0$\\
			
			&$f(Q)_{HT1}$	&$-$&$-$&$-$&$985.1$&$991.1$&$1005.8$&$3.2$&$8.1$  \\
			
			&$f(Q)_{HT2}$	&$-$&$-$&$-$&$984.2$&$990.2$&$1004.9$&$2.3$&$7.2$\\
			
			\hline
		\end{tabular}		
		\label{tab1}		
	\end{center}	
\end{table*}



\
\begin{figure*}[!htb]
	\centering
	\includegraphics[width=8.65cm]{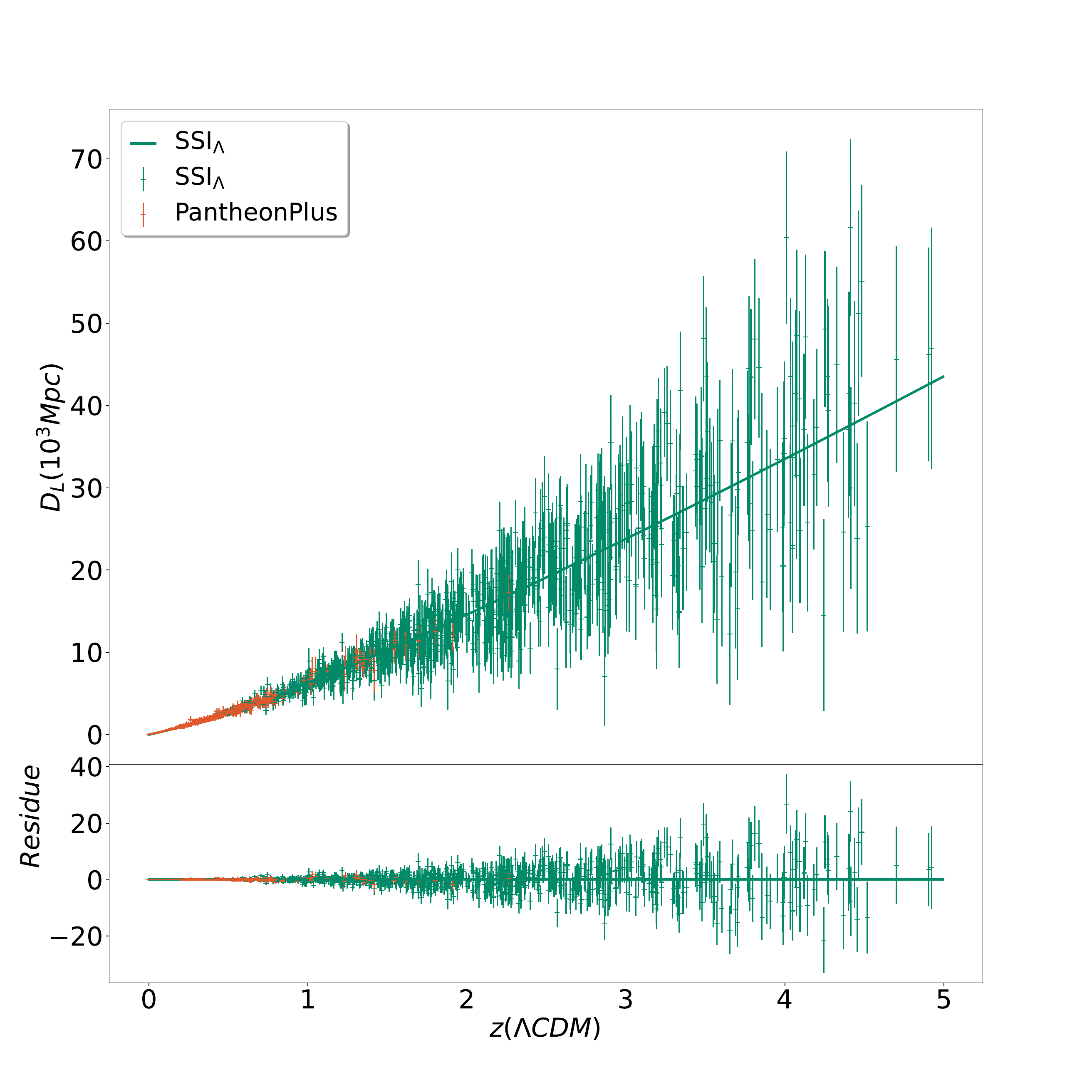}
	\includegraphics[width=8.65cm]{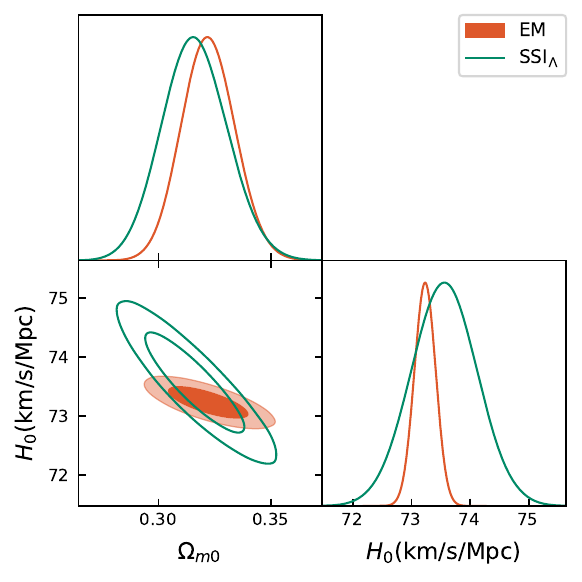}
	\caption{In the  left panel,  the $D_{L}$s of real and simulated data are compared. The red one with error-bar are from the  PantheonPlus combination; the green one with error-bar are from simulated  SS\Romannum{1}$_{\Lambda}$ data; and the  green line denotes the evolution of $D_{L}$ following the assumption of  SS\Romannum{1}$_{\Lambda}$ data which are based on $\Lambda$CDM model with its   best fitted EM values ($\Omega_{m0}=0.322$ and  $H_0=73.23$). In the right panel, the probability density functions (pdfs) with its 1$\sigma$ and 2$\sigma$ confidence regions for the parameters of $\Lambda$CDM model ($\Omega_{m0}$ and $H_0$) are shown.}
	\label{lcdm}
\end{figure*}


\begin{figure*}[!htb]
	\centering
	\includegraphics[width=8.65cm]{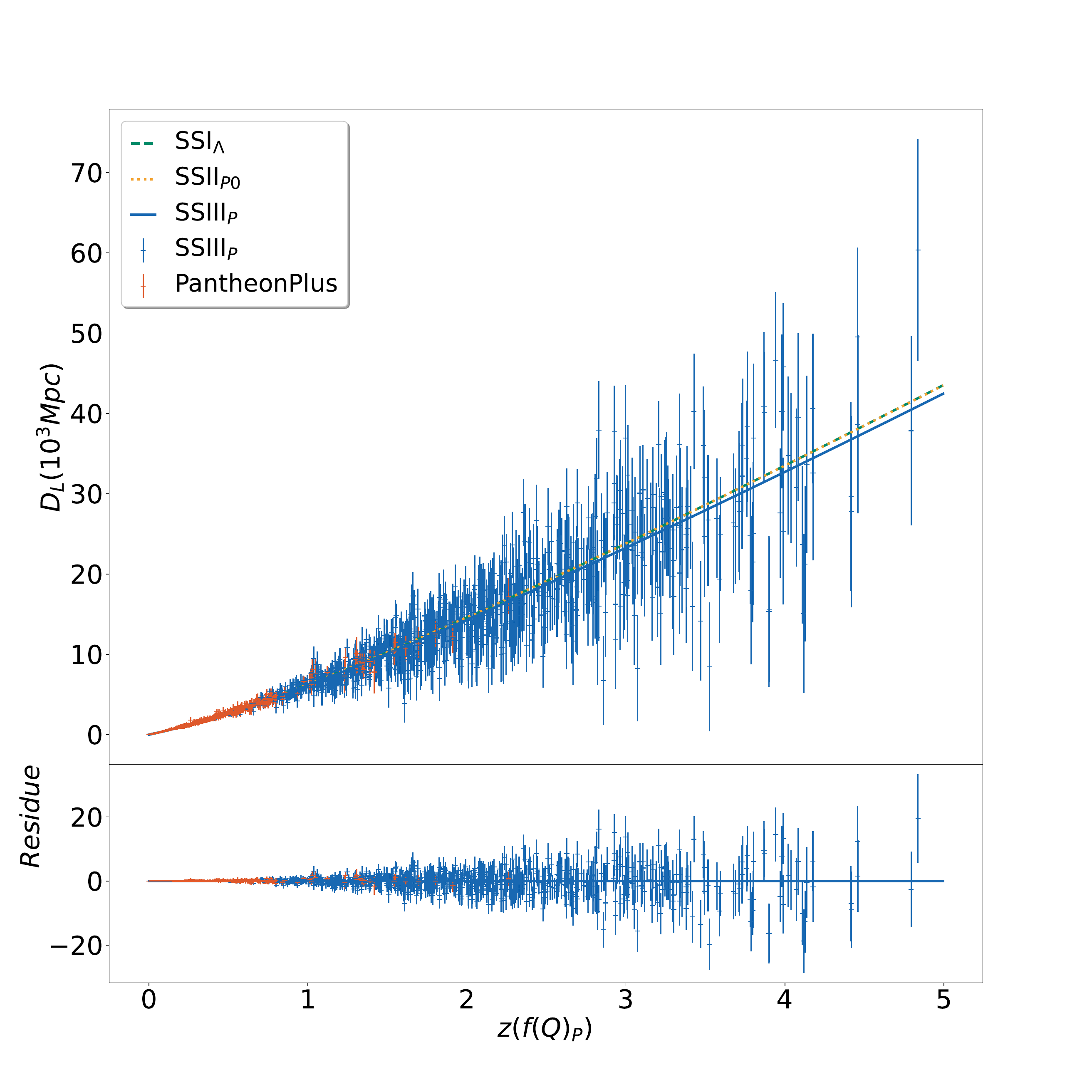}
	\includegraphics[width=8.65cm]{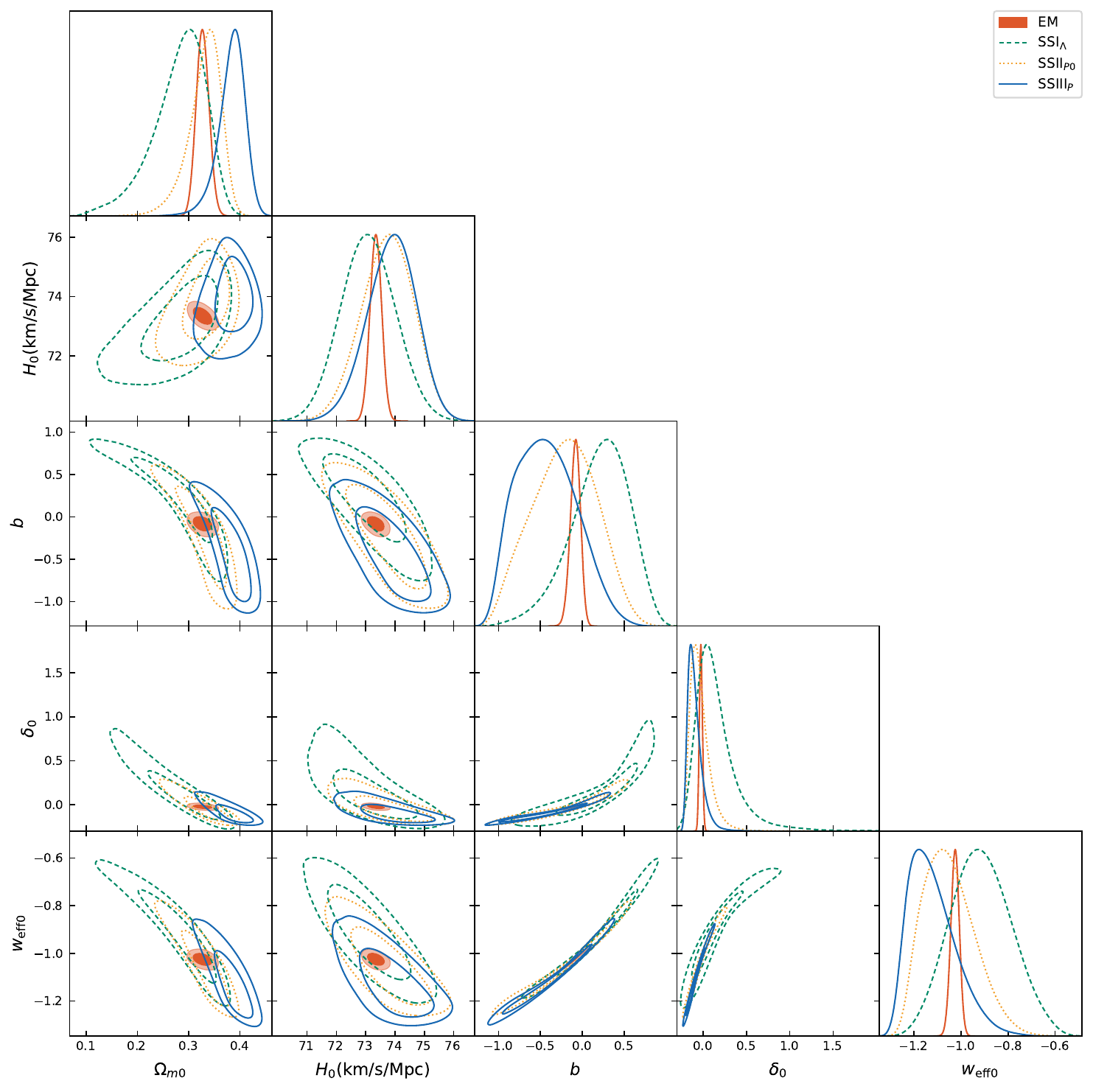}\\
	\includegraphics[width=8.65cm]{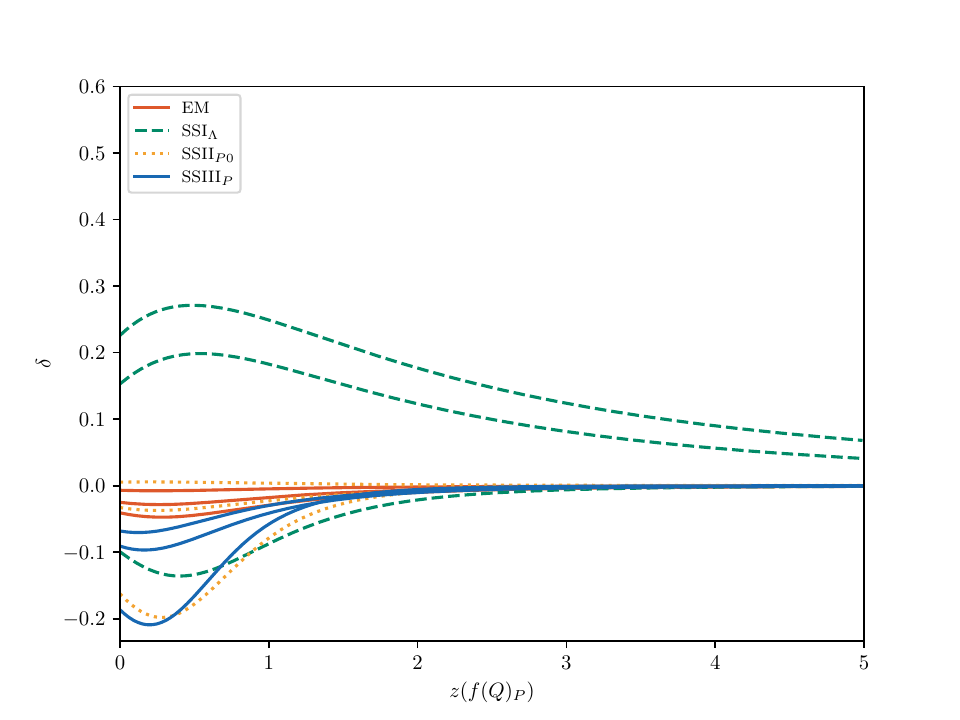}
	\includegraphics[width=8.65cm]{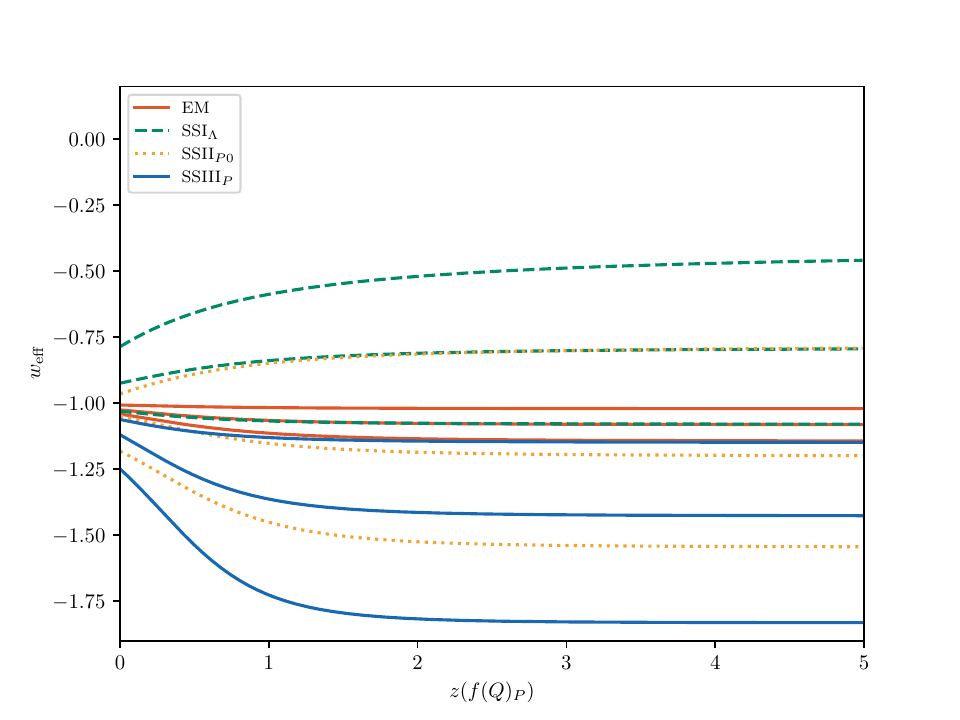}
	\caption{  The two upper panels are similar as Fig.\ref{lcdm}. In the upper left panel,  the $D_{L}$s of real and simulated data are compared. The red one with error-bar denote the PantheonPlus data as Fig.\ref{lcdm}. And we change the green solid line for SS\Romannum{1}$_{\Lambda}$ in Fig.\ref{lcdm} to the dashed green one for comparison. Additionally, the  orange line denotes the simulated   SS\Romannum{2}$_{P0}$ data with the best fitted EM values of $f(Q)_P$ model ($\Omega_{m0}=0.327$, $H_0=73.35$, $b=-0.082$ and $\delta_{0}=0$). The blue ones with error-bar  are   SS\Romannum{3}$_P$ data which has the same baseline values of $\Omega_{m0}$, $H_0$ and $b$ as SS\Romannum{2}$_{P0}$ simulation, but has a non-zero friction term ($\delta_{0}=-0.024$). In the upper right panel, the probability density functions with its 1$\sigma$ and 2$\sigma$ confidence regions for the parameters of $f(Q)_P$ model ($\Omega_{m0}$, $H_0$, $b$, $\delta_{0}$ and $w_{\mathrm{eff0}}$) are shown.
	And in  the  two bottom panels,  the evolutions of  $\delta$ and $w_{\mathrm{eff}}$ within 1$\sigma$ confidence intervals for the $f(Q)_{P}$ model under the constraints of EM and SS related data are shown.}
	\label{fp}
\end{figure*}
\begin{figure*}[!htb]
	\centering
	\includegraphics[width=8.65cm]{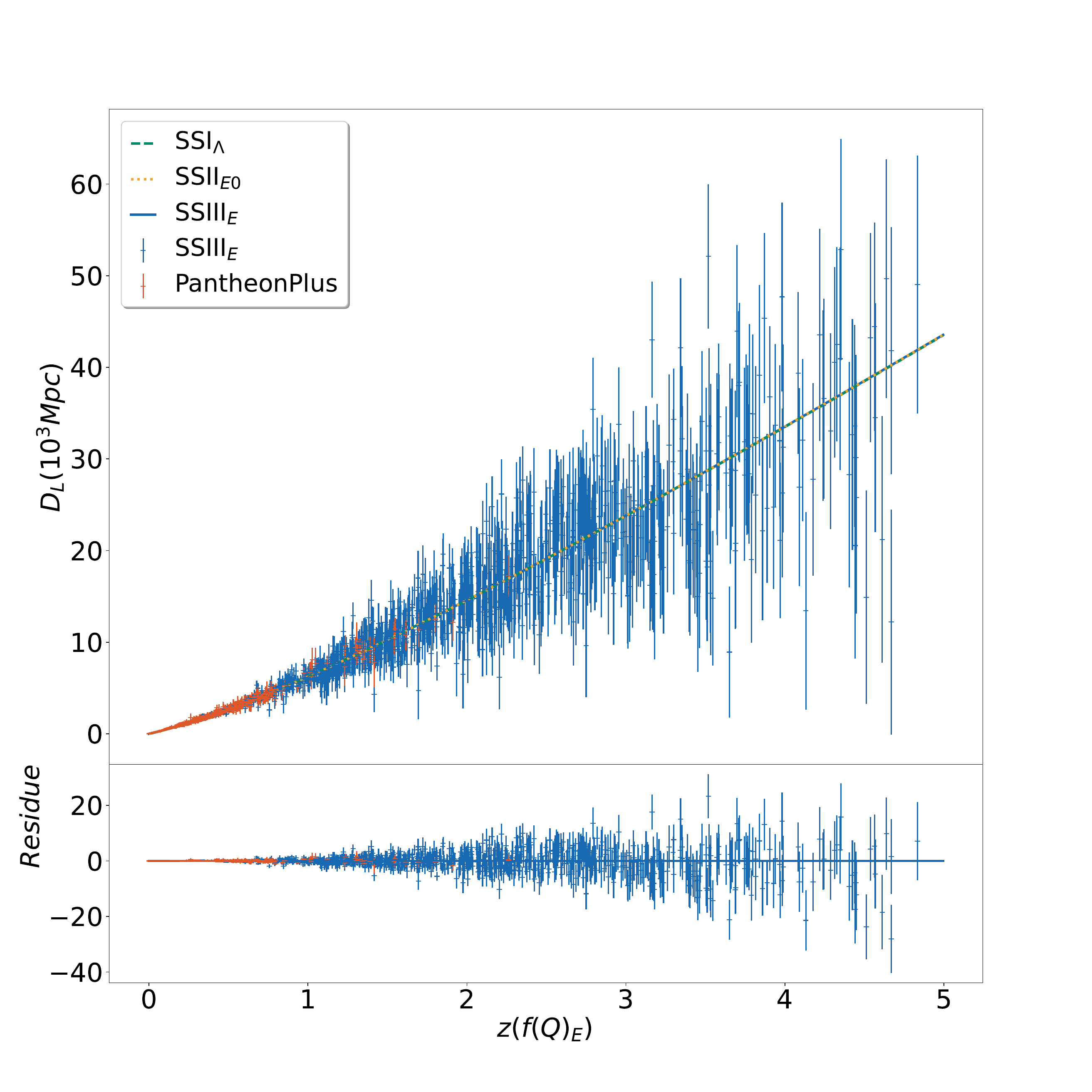}
	\includegraphics[width=8.65cm]{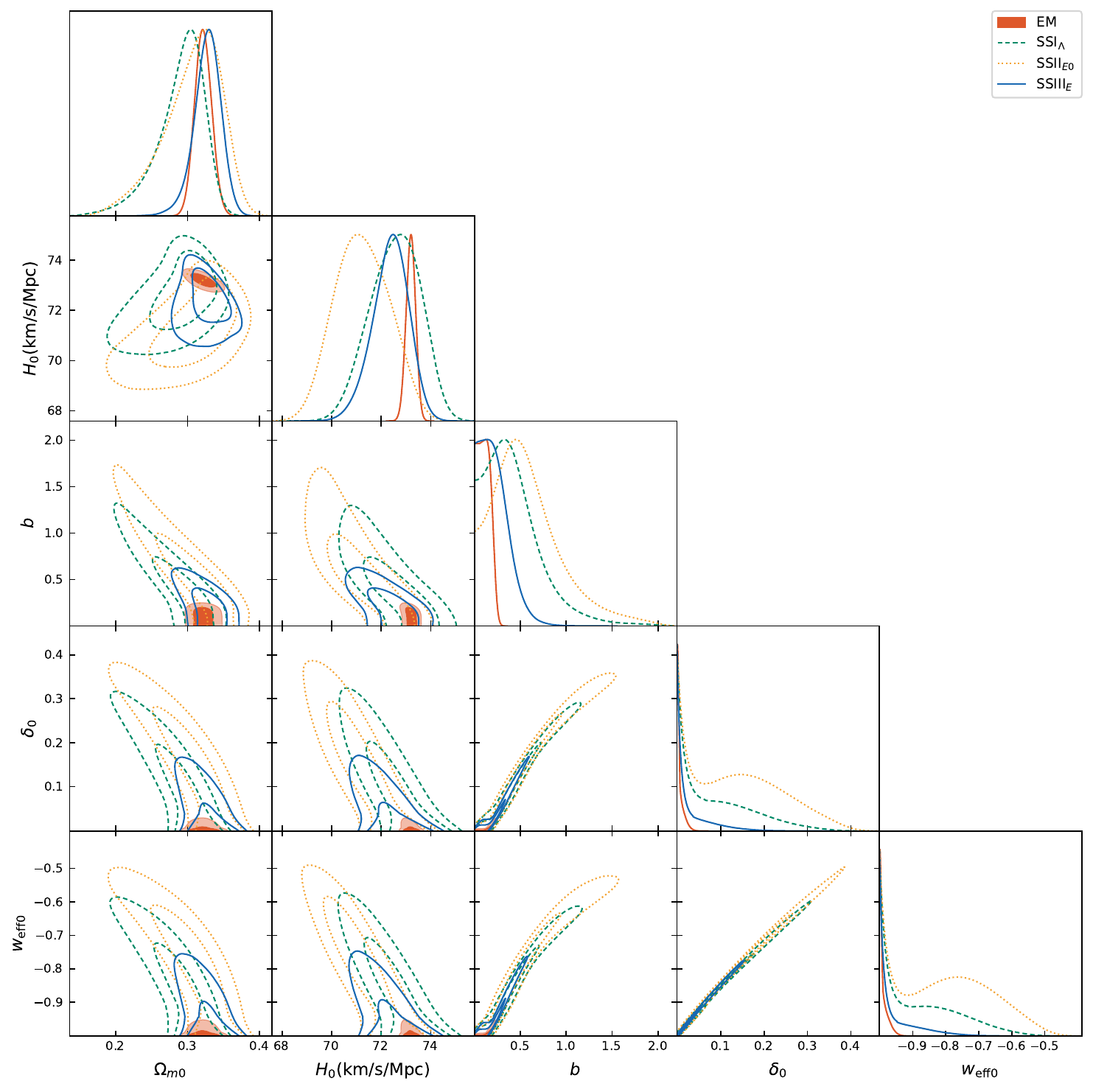}\\
	\includegraphics[width=8.65cm]{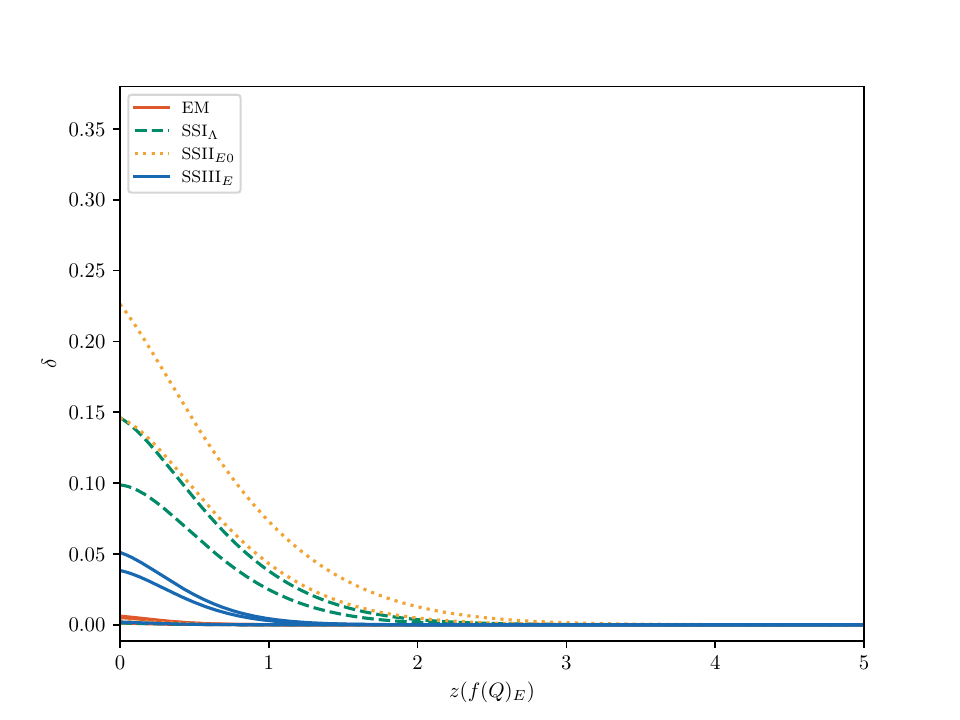}
	\includegraphics[width=8.65cm]{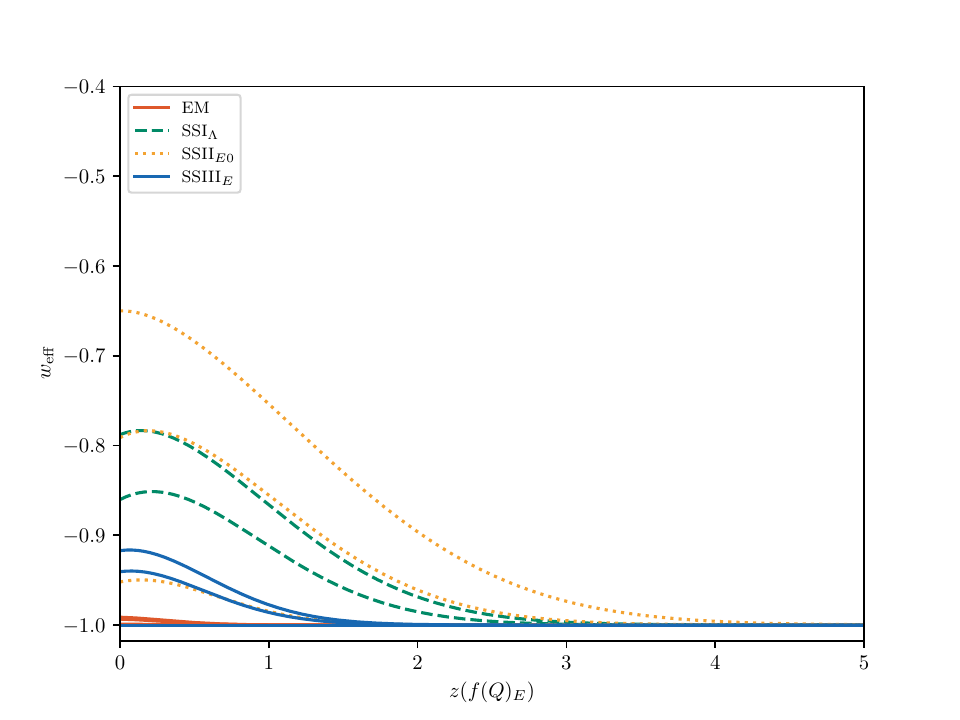}	
	\caption{ The fiducial values in the $f(Q)_E$ simulations are  $\Omega_{m0}=0.322$, $H_0=73.19$ and $b=0.106$ with $\delta_{0}=0$ for SS\Romannum{2}$_{E0}$ and $\delta_{0}=0.005$ for SS\Romannum{3}$_E$. The others are the same as Fig.\ref{fp}.}
	\label{fe}
\end{figure*}


\section{Results and discussion}\label{results}
To see the precision values, the constraining best  fitted parameter values with $1\sigma$ and $2\sigma$ standard errors are presented  in Table \ref{tab}. 
We also list the $\chi^2$, $AIC$ and $BIC$ results in Table \ref{tab1} as supplements.   The main results of $\Lambda$CDM and $f(Q)$ cosmologies are presented in Figs. \ref{lcdm},  \ref{fp},  \ref{fe}, \ref{fpe},   \ref{fht1} and  \ref{fht2}  respectively. Explicitly, the data comparisons, the  parameter probability density functions (pdfs) with its 1$\sigma$ and 2$\sigma$ confidence regions, and the evolutions of  $\delta$ and $w_{\mathrm{eff}}$ within 1$\sigma$ confidence intervals are plotted here.  

As the data comparison figures shows, the evolution values of $D_L^{SS}$ in $f(Q)_E$ model are un-distinguished. And the evolution values of $D_L^{SS\Romannum{1}_\Lambda}$ are un-distinguished with that of $D_L^{SS\Romannum{2}}$ in the $f(Q)_P$ and $f(Q)_{HT2}$ models. Meanwhile, the evolution values of $D_L^{SS\Romannum{1}_\Lambda}$ are slightly smaller than that of $D_L^{SS\Romannum{2}}$ in the $f(Q)_{HT1}$ model, but slightly larger than that of $D_L^{SS\Romannum{2}}$ in the $f(Q)_{PE}$ model. 
And the evolution values of $D_L$ related to SS\Romannum{3} simulation  are smaller than that  of  SS\Romannum{1}$_\Lambda$ and SS\Romannum{2} simulations. Explicitly,  for SS\Romannum{2} simulations, $D_L^{SS\Romannum{2}_{PE}}<  D_L^{SS\Romannum{2}_{P}}\simeq D_L^{SS\Romannum{2}_{E}}\simeq D_L^{SS\Romannum{2}_{HT2}}\simeq D_L^{ SS\Romannum{1}_{\Lambda}}< D_L^{SS\Romannum{2}_{HT1}}$; for SS\Romannum{3} simulations, $D_L^{SS\Romannum{3}_{HT2}}\ll  D_L^{SS\Romannum{3}_{HT1}}\simeq D_L^{SS\Romannum{3}_{PE}}<D_L^{SS\Romannum{3}_{P}}< D_L^{SS\Romannum{3}_{E}}\simeq D_L^{ SS\Romannum{1}_{\Lambda}}$.

And as the triangle plots show, the contours of $f(Q)_E$ model are not closed because of the prior $b>0$. While in the other models, the contours are smoothly closed and the probability density functions (pdfs) are Gaussian-distributed.
Roughly, for all the models, the   tightest constraints come from the EM data. Especially, for the $\Lambda$CDM, $f(Q)_E$ and $f(Q)_{HT2}$  models,  $\Omega_{m0}$ is around $0.322$ with  $0.024$ ($0.045$) as $1\sigma$ ($2\sigma$)  error range. In the other models, the  error ranges are similar but the best fitted $\Omega_{m0}$s sightly  shift.   Furthermore, the error ranges of $H_0$ in EM constraints are similar. In another saying, all the simulations are based on similar $\Omega_{m0}$ and $H_0$.  
The Hubble tension could be alleviated to $0.07\sigma$ level under the constraint of EM in $\Lambda$CDM model. Then, it is not surprising that the Hubble tension could be alleviated in other constraints except that of  SS\Romannum{3}$_{HT2}$.  Surprisingly, the simulation of SS\Romannum{3}$_{HT2}$ is problematical where the $\Omega_{m0}$ is very closed to its upper limit with  $12.56\sigma$ Hubble tension. 

In this discussion, the  $\Lambda$CDM  model is used as baseline.  	
As Fig. \ref{lcdm}  shows, the $\Omega_{m0}-H_0$ contour of EM is much smaller than that of the SS\Romannum{1}$_{\Lambda}$ data  in the $\Lambda$CDM model. The direction of $\Omega_{m0}-H_0$ contour is changed slightly as well. 
In the next, we discuss all the $f(Q)$ models one by one. 

\subsection{Discussion on the $f(Q)_{P}$  model}
As Table \ref{tab} shows, by comparing the results from SS\Romannum{1}$_{\Lambda}$  and SS\Romannum{2}$_{P0}$, the model effect brings a large shift of the best fitted value of $ \Omega_{m0}$ which is $\Delta \Omega_{m0}=0.049$ in $f(Q)_{P}$ model. And by comparing SS\Romannum{2}$_{P0}$ and SS\Romannum{3}$_{P}$ constraints, the extra friction term brings  shift of the best fitted value of $ \Omega_{m0}$ as $\Delta \Omega_{m0}=0.053$ which is $221\%$ of the $1\sigma$ regime of EM constraint ($\Delta\Omega_{m0}^{1\sigma}=0.024$).
The model effects are comparable with  the extra friction term effects in the $f(Q)_P$ model.

As Fig.\ref{fp} shows, the constraining tendencies of SS related data are similar. And their contours  are overlapped with the EM ones in $2\sigma$ ranges. Especially, the $\Omega_{m0}$ related to SS\Romannum{3}$_{P}$ is as large as $0.383^{+0.028+0.047}_{-0.018-0.053}$. The correlation  of $\Omega_{m0}-H_{0}$ contour is negative in the EM constraint while it is positive in the SS related constraints. This phenomenon could help to break the degeneration between parameters.

Furthermore, the $b=0$, $\delta=0$ and  $w_\mathrm{{eff}}=-1$ are not included in $1\sigma$ range of EM and SS\Romannum{3}$_{P}$ data for  $f(Q)_{P}$ model.
In another saying, the   $f(Q)_{P}$ model could be distinguished from $\Lambda$CDM  in $1\sigma$ ranges.
And at high $z$, the deviations from $\Lambda$CDM model  become evident if $w_\mathrm{{eff}}\neq-1$, while most $\delta$s tends to $0$ which correspond to a flat $w_\mathrm{{eff}}$.

As Table \ref{tab1} shows, the $\chi^{2}_{CC}$  and $\chi^{2}_{DESI}$ of $f(Q)_P$ model are smaller than that of $\Lambda$CDM model. While, the $\chi^{2}_{PantheonPlus}$ of $f(Q)_P$ model is larger than that of $\Lambda$CDM model. After EM data combination, $\Delta AIC_{EM}=2.3$, the $f(Q)_P$ model could be regarded as equal as $\Lambda$CDM model. But  $\Delta BIC_{EM}= 7.8$ denotes the $f(Q)_P$ model is ``punished" by the EM data. And in the SS\Romannum{1}$_\Lambda$ constraint, $\Delta AIC_{SS\Romannum{1}_\Lambda}=2.2$ and $\Delta BIC_{SS\Romannum{1}_\Lambda}=7.1$ which are similar to the EM case.

\subsection{Discussion on the $f(Q)_{E}$ model} 
As Table \ref{tab} shows, by comparing the results from SS\Romannum{1}$_{\Lambda}$  and SS\Romannum{2}$_{E0}$, the model effect brings a shift of the best fitted value of $ \Omega_{m0}$ which is $\Delta \Omega_{m0}=0.017$ in $f(Q)_{E}$ model. While the extra friction term brings  shift of the best fitted value of $ \Omega_{m0}$ as $\Delta \Omega_{m0}=0.021$ by comparing SS\Romannum{2}$_{E0}$ and SS\Romannum{3}$_{E}$. The $ \Omega_{m0}$ shifts of $f(Q)_{E}$ model are comparable in the model and extra friction effects.

As the $f(Q)_P$ model, the shapes of SS related data are similar and the smallest contours are still from the SS\Romannum{3}$_{E}$ simulation. Comparing with the EM and SS related data, the direction of $\Omega_{m0}-H_{0}$ contour is changed. 
Specially, the Hubble tension from EM constraint is as small as $0.02\sigma$ which is smallest among all constraints.

And we obtain  a positive $\delta_{0}$ and a quintessence-like $w_\mathrm{{eff}}$ in all $f(Q)_{E}$ constraints.
The evolutions of $w_\mathrm{{eff}}$ and $\delta$, which are quite flat, almost follow $\Lambda$CDM model at $z>2$. In the other saying, the deviations from $\Lambda$CDM model become ignored at high $z$.
	
In contrast to the $f(Q)_{P}$ model, the $\chi^{2}_{DESI}$ of $f(Q)_E$ model is larger than that of $\Lambda$CDM model, while the $\chi^{2}_{CC}$ is equal to that of $\Lambda$CDM model. And the $\chi^{2}_{PantheonPlus}$ of $f(Q)_E$ model is smaller than that of $\Lambda$CDM model. Finally after combination of EM data, the information criterion gives similar results to $f(Q)_{E}$ model as $f(Q)_{P}$ model. Explicitly, $\Delta AIC_{EM}=3.4$ shows the $f(Q)_{E}$ model could be regarded as equal as $\Lambda$CDM model. And  $\Delta BIC_{EM}= 8.9$ which denotes the $f(Q)_E$ model is ``punished" by the EM data. Furthermore in the SS\Romannum{1}$_\Lambda$ constraint, $\Delta AIC_{SS\Romannum{1}_\Lambda}=2.4$ and $\Delta BIC_{SS\Romannum{1}_\Lambda}=7.3$ which are similar to the EM case as well.

\subsection{Short summary on the $\Lambda$CDM-like models}
Generally, both the $f(Q)_{P}$ and $f(Q)_{E}$ models give out small Hubble tensions as $\Lambda$CDM model. And the $1\sigma$ ($2\sigma$) constraining regions of all the  SS related data  of $f(Q)_{P}$($f(Q)_{E}$) model are much larger than that of EM data. Especially, the best fitted $\Omega_{m0}$ of SS\Romannum{1}$_\Lambda$ are much smaller than that of other data while the plots of $D_L$ are un-distinguishable. This phenomenon hints the model effect, which is comparable with the friction term,  could not be ignored in the $\Lambda$CDM-like model simulations.

In the EM and SS\Romannum{1}$_\Lambda$ data, the $f(Q)_{P}$ and $f(Q)_{E}$  models  are favored by AIC, but  ``punished" by BIC.
Theoretically, the $b$ parameter may effect cosmic perturbations giving an intriguing division between background and perturbation behavior in terms of model parameters \cite{Capozziello:2023giq,Chen:2010va,Mhamdi:2024lxe}. In the future, the growth factor data which are derived from matter perturbations could be used to give out more information.
 

\begin{figure*}[!htb]
	\centering
	\includegraphics[width=8.6cm]{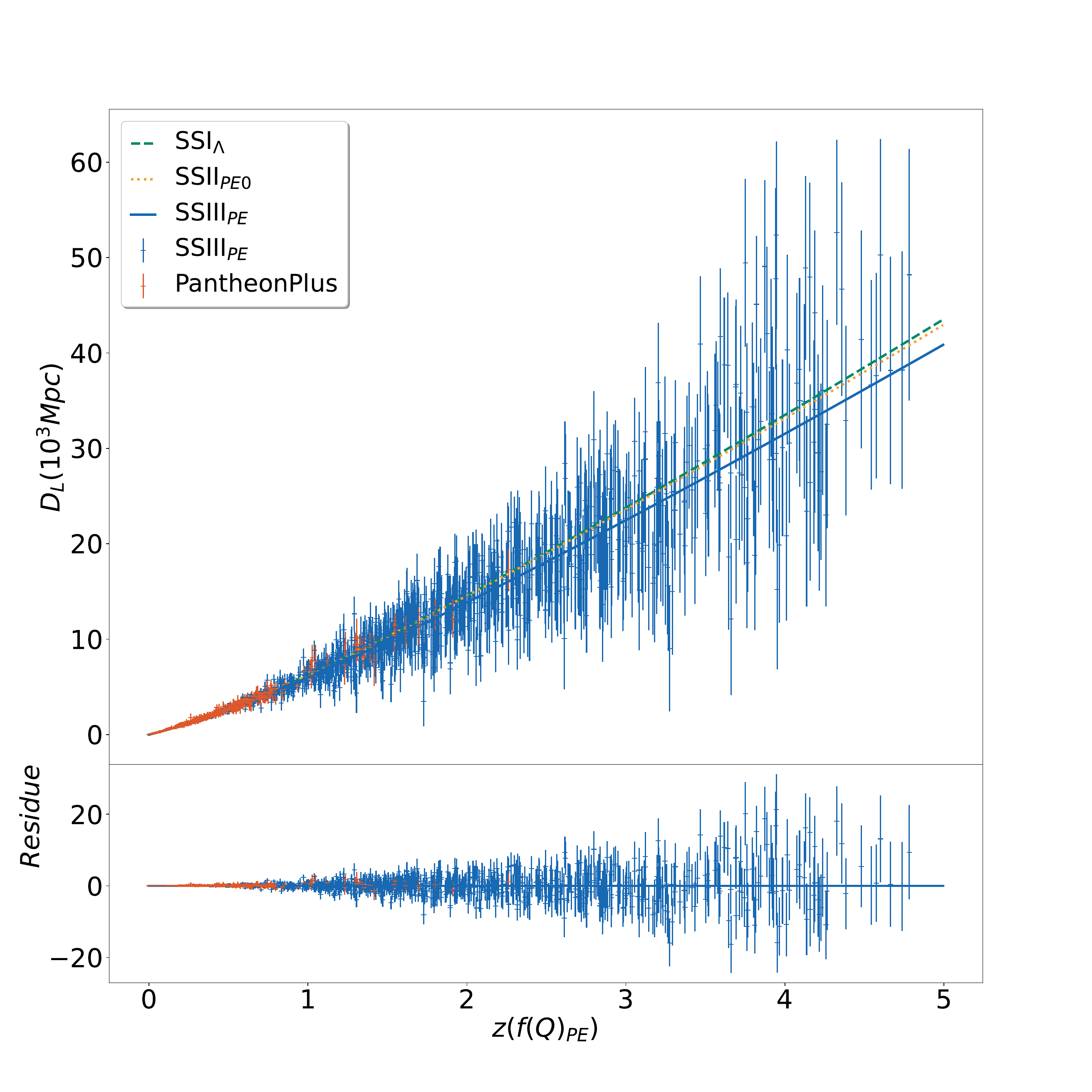}
	\includegraphics[width=8.6cm]{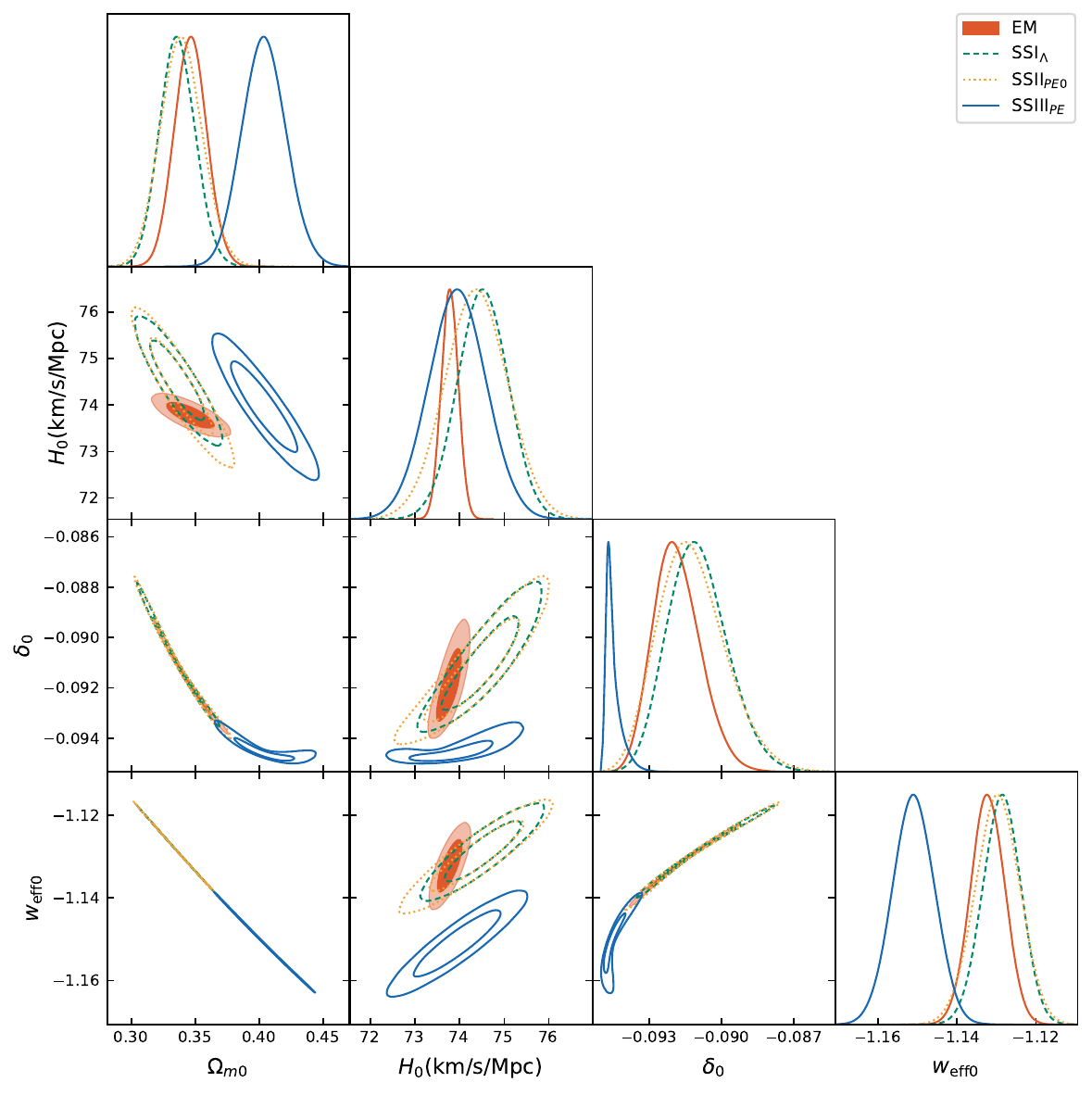}\\
	\includegraphics[width=8.6cm]{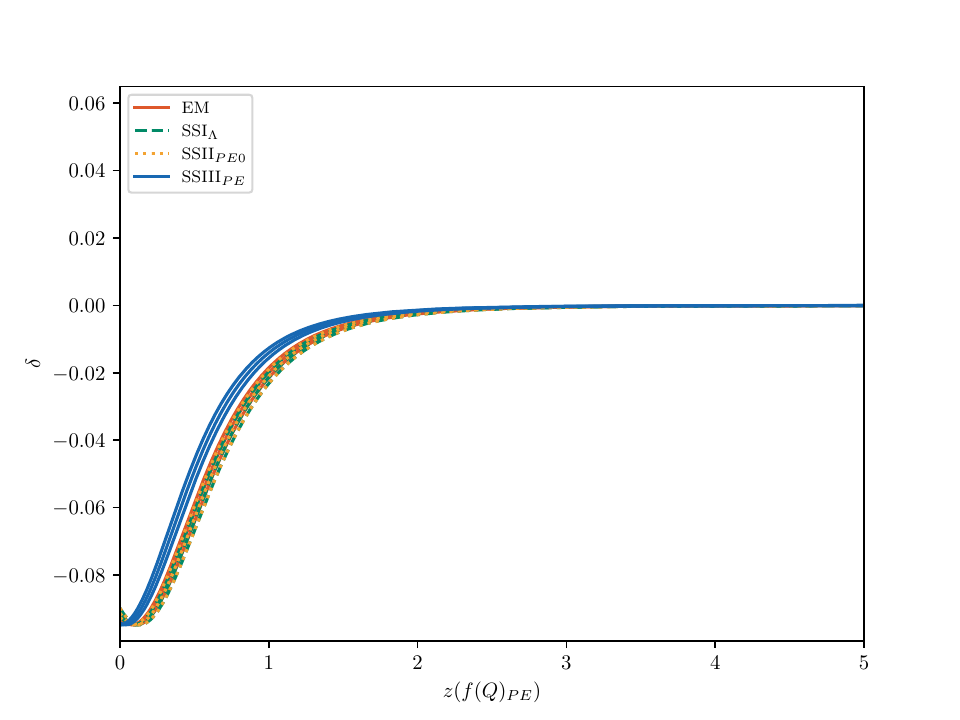}
	\includegraphics[width=8.6cm]{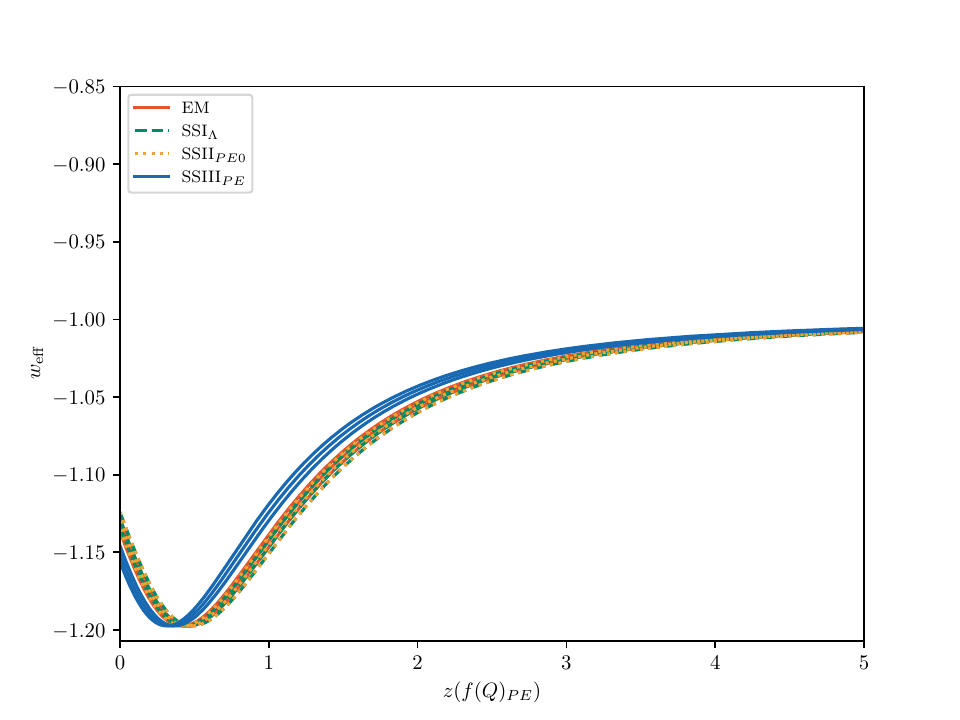}
	\caption{The fiducial values in the $f(Q)_{PE}$ simulations are  $\Omega_{m0}=0.347$ and $H_0=73.78$  with $\delta_{0}=0$ for SS\Romannum{2}$_{PE0}$ and $\delta_{0}=-0.092$ for SS\Romannum{3}$_{PE}$. The others are the same as Fig.\ref{fp}.}
	\label{fpe}
\end{figure*}
\begin{figure*}[!htb]
	\centering
	\includegraphics[width=8.65cm]{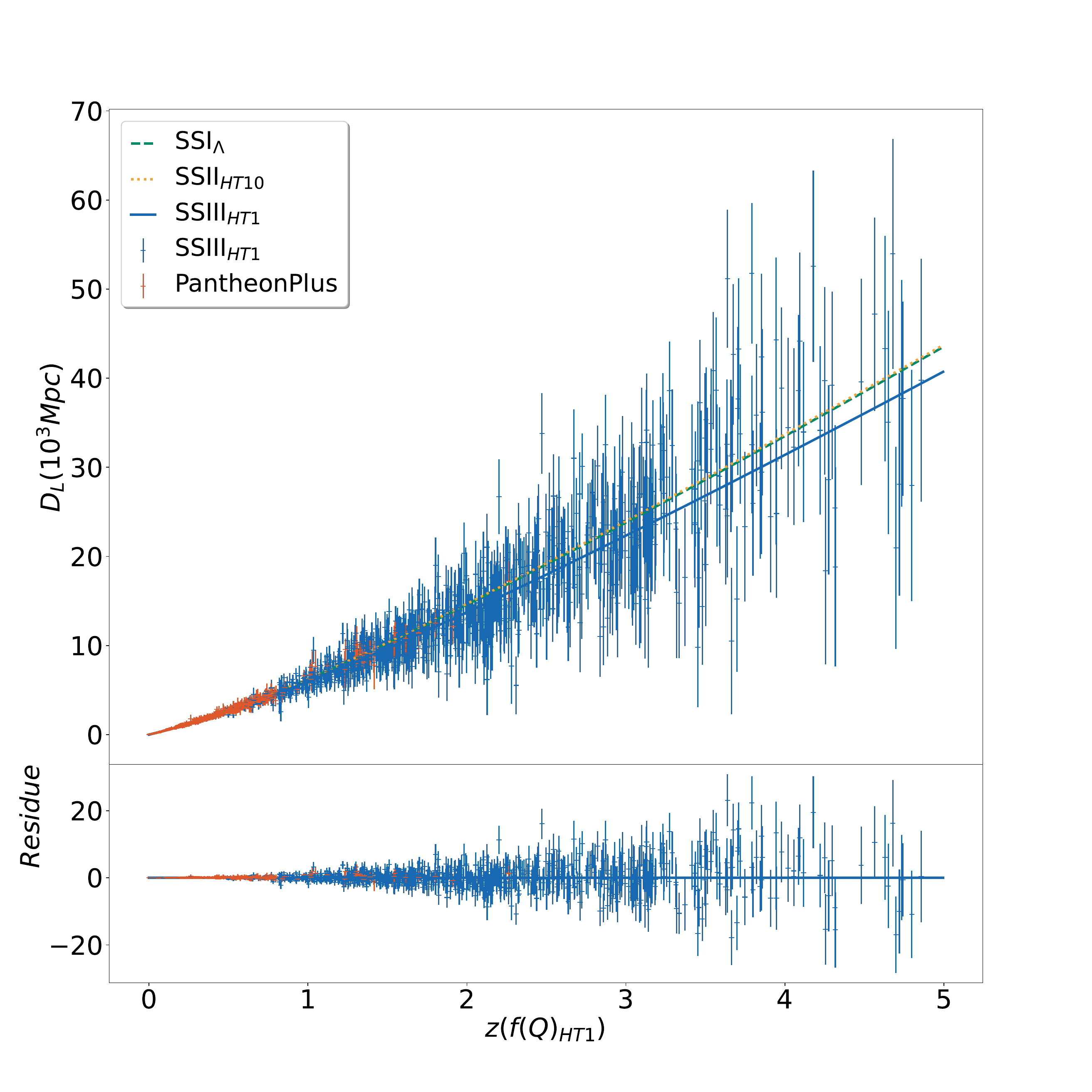}
		\includegraphics[width=8.65cm]{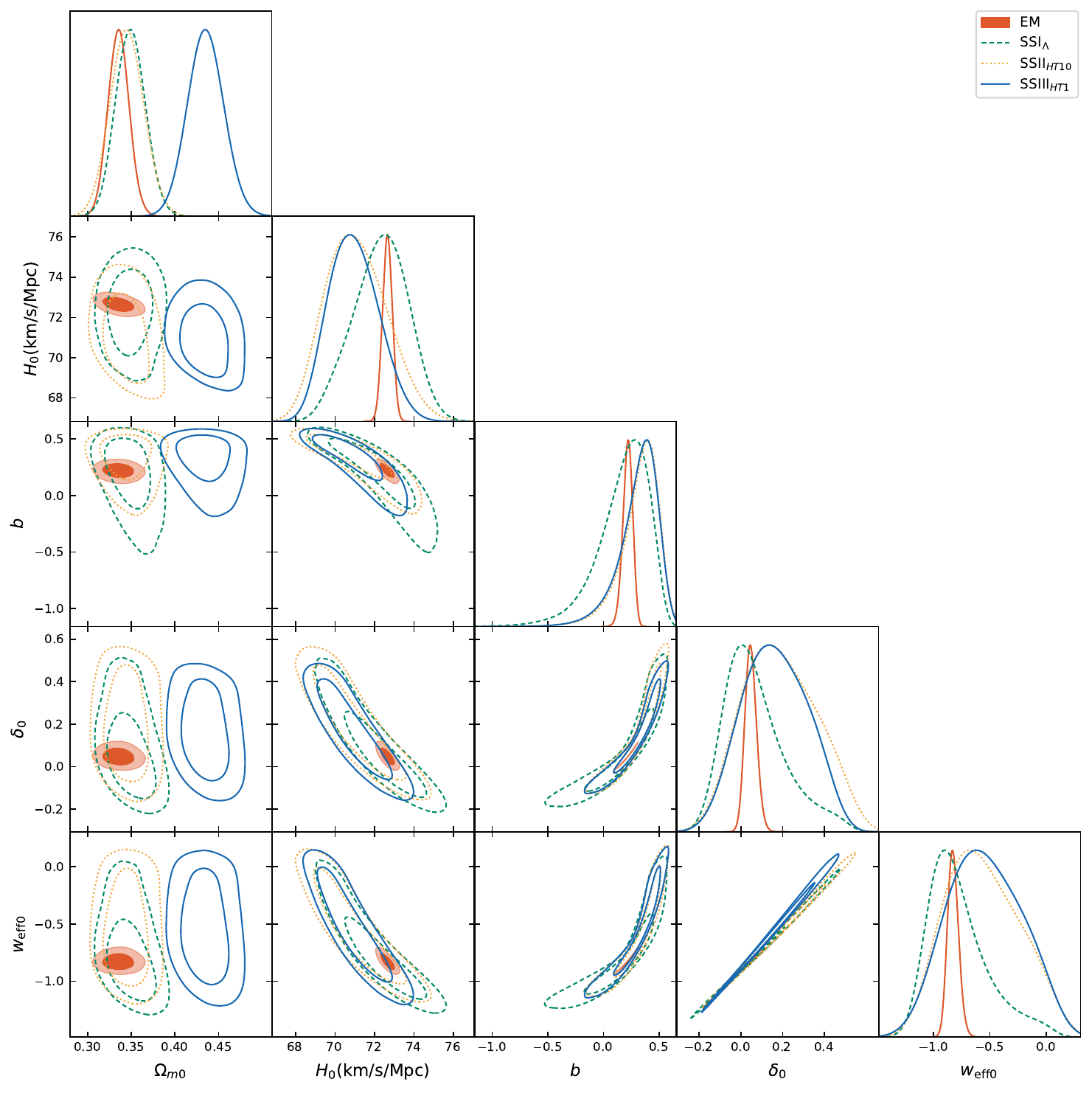}	\\
		\includegraphics[width=8.65cm]{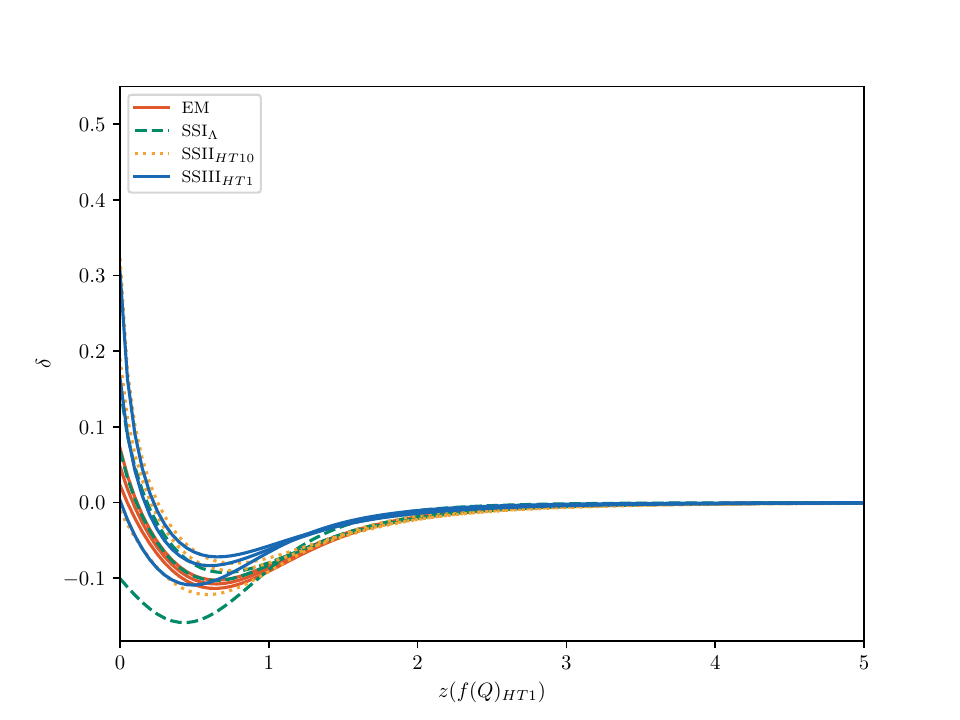}
		\includegraphics[width=8.65cm]{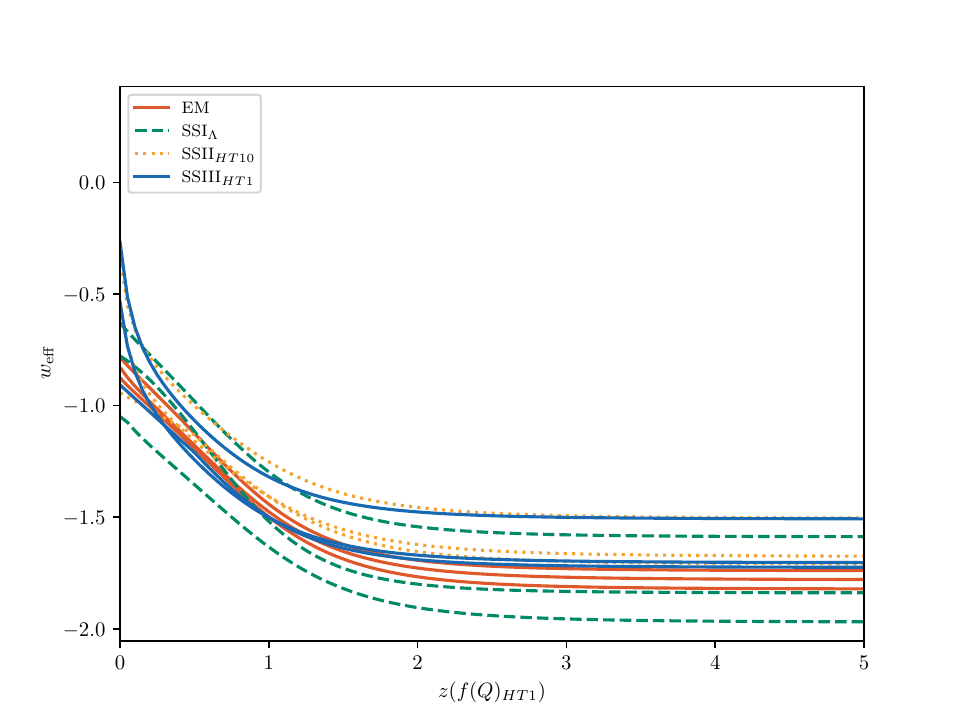}
	\caption{The fiducial values in the $f(Q)_{HT1}$ simulations are  $\Omega_{m0}=0.336$, $H_0=72.65$ and $b=0.218$  with $\delta_{0}=0$ for SS\Romannum{2}$_{HT10}$ and $\delta_{0}=0.048$ for SS\Romannum{3}$_{HT1}$. The others are the same as Fig.\ref{fp}.}
	\label{fht1}
\end{figure*}

\begin{figure*}[!htb]
	\centering
\includegraphics[width=5.75cm]{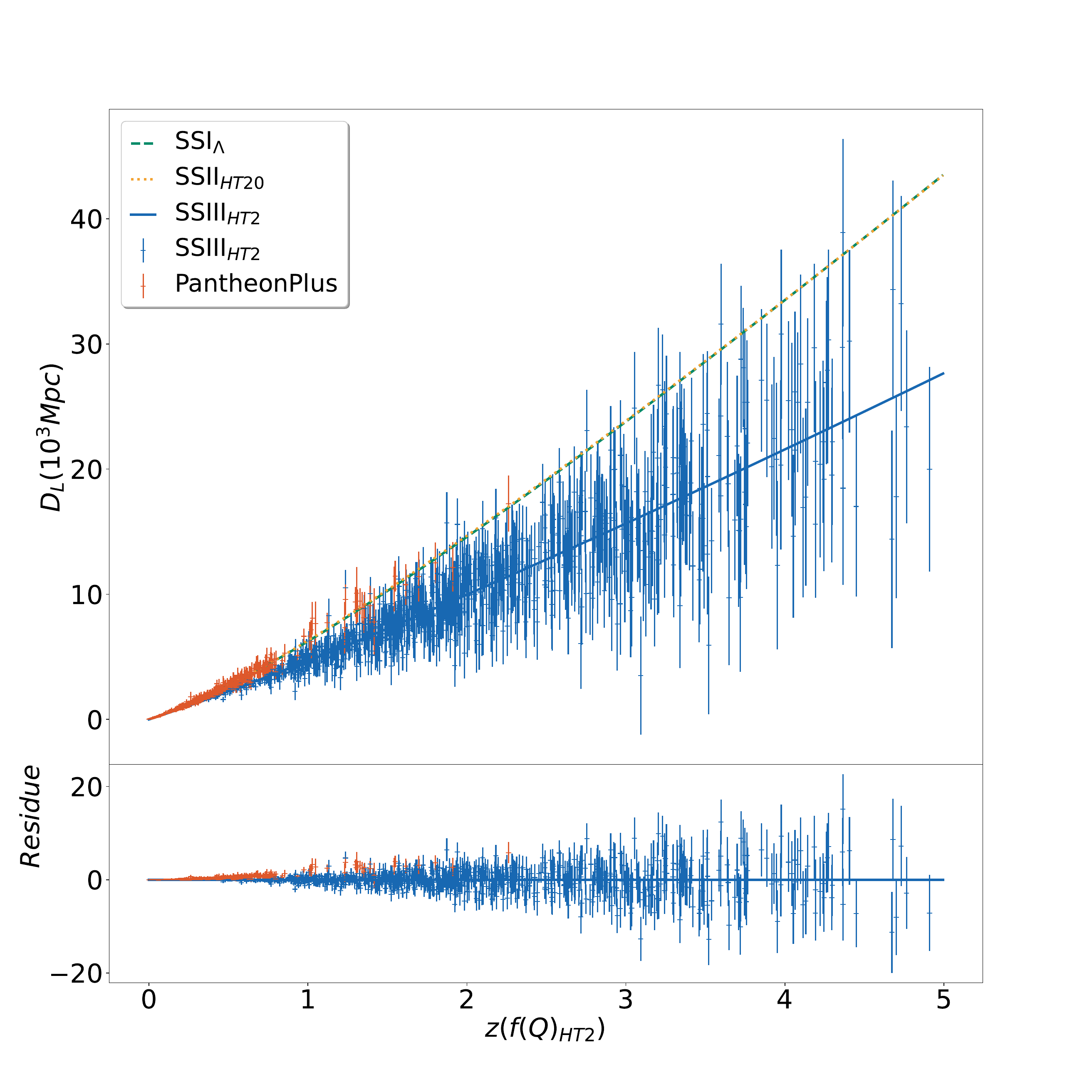}
\includegraphics[width=5.75cm]{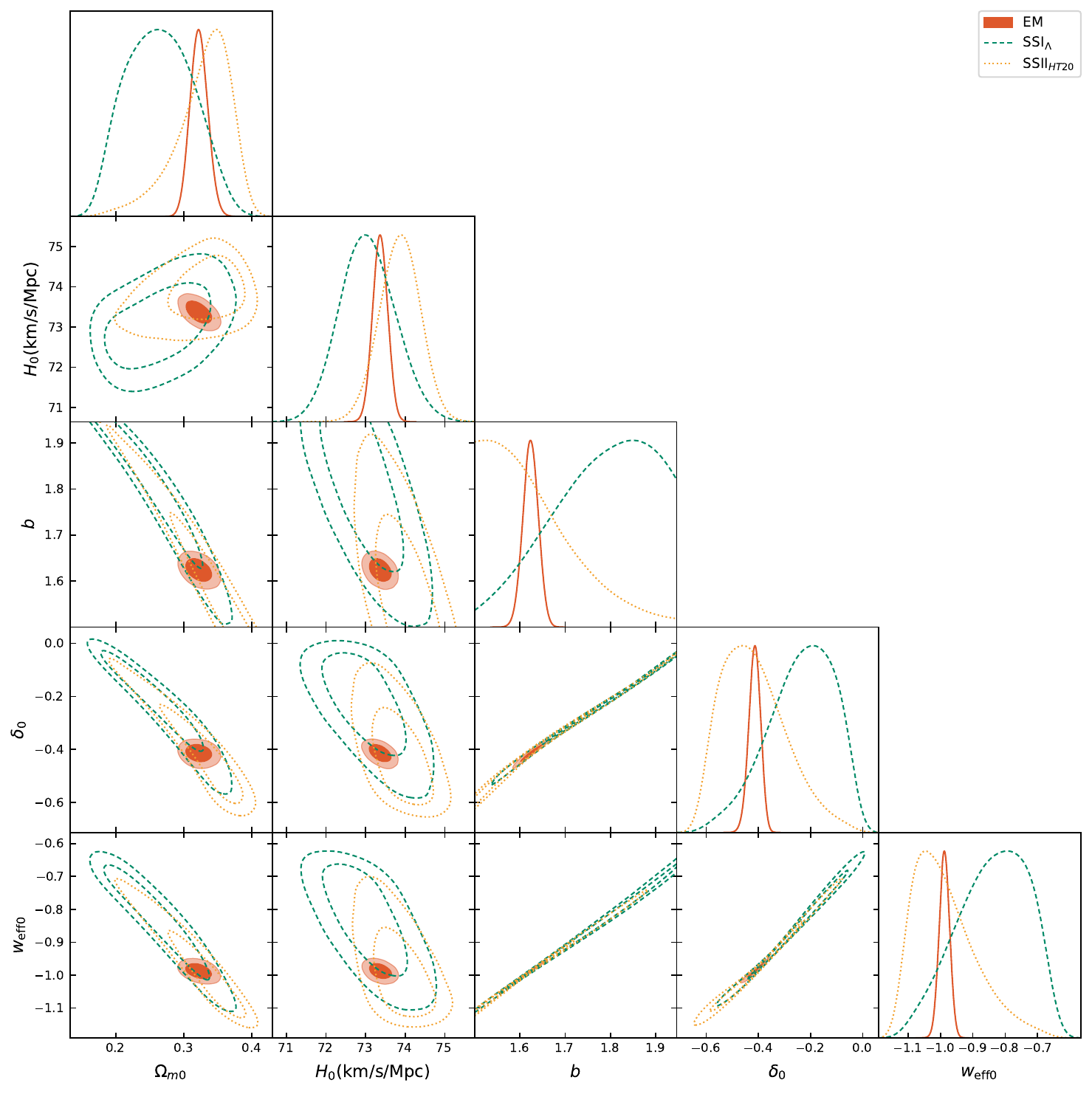}	
\includegraphics[width=5.75cm]{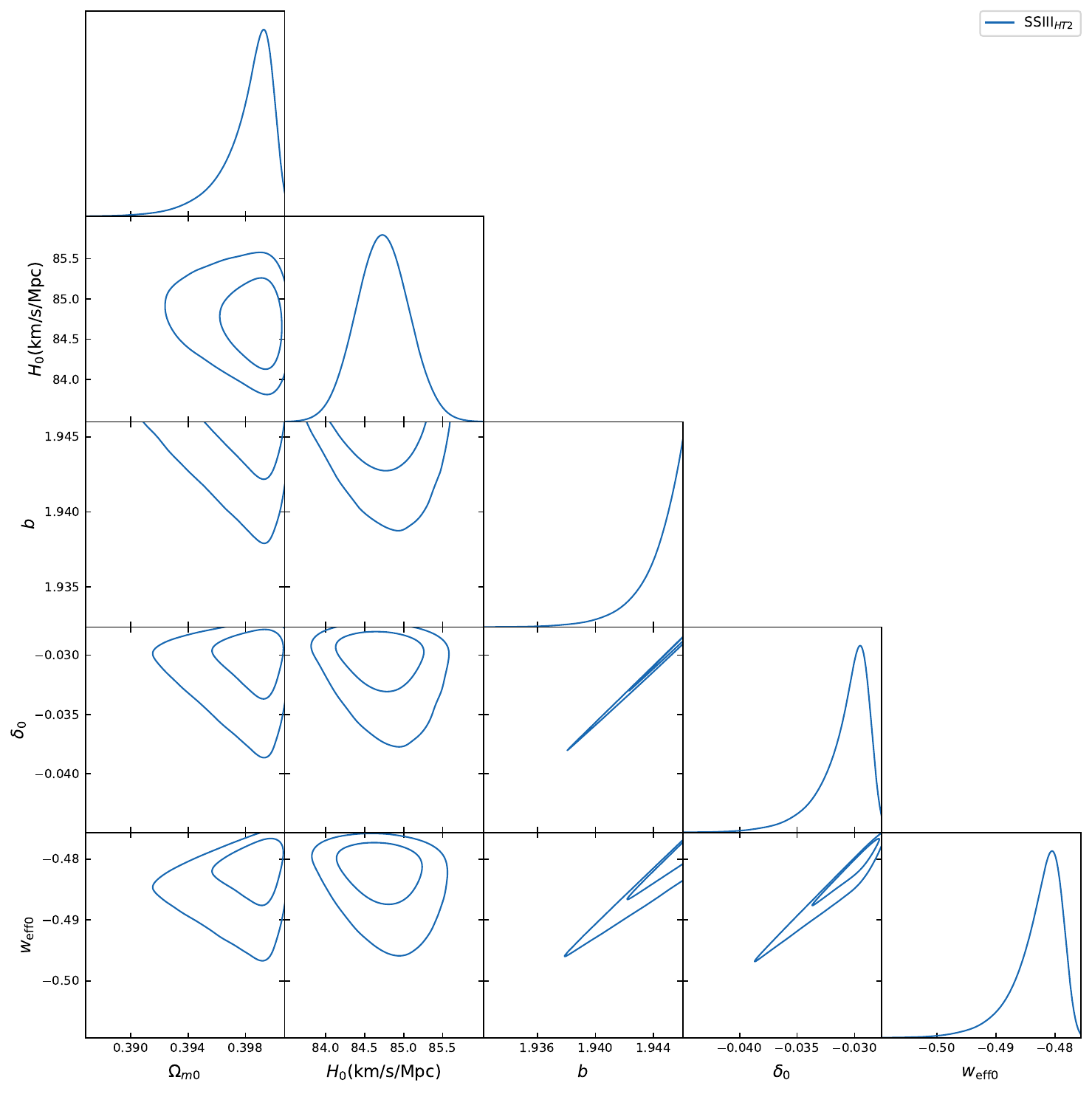}
	\caption{Here, as the constraining result of $\Omega_{m0}$ of SS\Romannum{3}$_{HT2}$ is very close to its upper prior, we assume $\Omega_{m0}<0.4$ for SS\Romannum{3}$_{HT2}$  and list the results at the right panel. The others are the same as Fig.\ref{fp}, except that the fiducial values in the $f(Q)_{HT2}$ SS simulations are  $\Omega_{m0}=0.322$, $H_0=73.37$ and $b=1.624$  with $\delta_{0}=0$ for SS\Romannum{2}$_{HT20}$ and $\delta_{0}=-0.415$ for SS\Romannum{3}$_{HT2}$.}
	\label{fht2}
\end{figure*}
\subsection{Discussion on the  $f(Q)_{PE}$ model}
The $f(Q)_{PE}$ model has the same free parameters with $\Lambda$CDM model. 
As Fig.\ref{fpe} shows, the $H_0$ related contours derived from the SS related data  are parallel. The  SS\Romannum{1}$_{\Lambda}$ and SS\Romannum{2}$_{PE0}$ related contours are closed to each other. And they have overlapped with the EM contours whose directions are changed. While the SS\Romannum{3}$_{PE}$ related contours are separated with the other data because of its $ \Omega_{m0}$ as large as $0.404^{+0.017+0.033}_{-0.017-0.031}$. 
The shapes of contours $\Omega_{m0}-w_\mathrm{{eff0}}$, $\Omega_{m0}-\delta_{0}$ and  $\delta_{0}-w_\mathrm{{eff0}}$  are narrow.
By comparing the results from SS\Romannum{1}$_{\Lambda}$  and SS\Romannum{2}$_{PE0}$,   the model effect brings a  slight shift of the best fitted value of $ \Omega_{m0}$ which is $\Delta \Omega_{m0}=0.003$. While the extra friction term brings a much larger shift of the best fitted value of $\Omega_{m0}$ which is $\Delta \Omega_{m0}=0.065$.

In the $f(Q)_{PE}$ model, as Table \ref{tab} and Fig.\ref{fpe} show, the parameter $\delta_0$  is always smaller  than $0$ in $2\sigma$ ranges. The best fitted values of $\delta$ and their $2\sigma$ regimes are around $-0.092$ for all the data. Precisely speaking, the $\delta_0=0$ is excluded in all the constraints of the $f(Q)_{PE}$ model where the extra friction term plays an important role in the simulations.  
All the constraint results exclude $w_{\rm eff0}=-1$ in $2\sigma$  regimes as well which means this model could be distinguished from the $\Lambda$CDM model.  
As $z$ increasing, the shapes of evolution of $w_\mathrm{{eff}}$ and $\delta$ are  similar which corresponds to the narrow positive correlation between $w_\mathrm{{eff}}$ and $\delta$. The value of $w_{\rm eff}$ gradually approaches $-1$, which mimics the standard $\Lambda$CDM model while it is still in the phantom range which did not cross $w_{\rm eff0}=-1$.   
And, $\delta$ gradually tends  to $0$ which  corresponds to $\Lambda$CDM model as well. These results are consistent with the analytic calculations in Eqs.(\ref{w1}) and (\ref{deltae}).

As Table \ref{tab1} shows, the $\chi^{2}_{CC}$ and $\chi^{2}_{DESI}$ of $f(Q)_{PE}$ model are larger than that of $\Lambda$CDM model. In the contrast, the difference between the $\chi^{2}_{PantheonPlus}$ of $f(Q)_{PE}$ model and that of $\Lambda$CDM model is as large as  $15.7$. After combination of EM data, $\Delta AIC_{EM}=22.6$ and  $\Delta BIC_{EM}= 22.6$ which denote the $f(Q)_{PE}$ model could be excluded. And in the SS\Romannum{1}$_\Lambda$ constraint, $\Delta AIC_{SS\Romannum{1}_\Lambda}=3.0$ and $\Delta BIC_{SS\Romannum{1}_\Lambda}=3.0$ are opposite to the EM constraint.


\subsection{Discussion on the $f(Q)_{HT1}$ model}
As Fig.\ref{fht1} shows, the parallel contours in $f(Q)_{HT1}$ model are the $\Omega_{m0}$ related ones which are derived from the SS related data.
Both the SS\Romannum{1}$_{\Lambda}$ and SS\Romannum{2}$_{HT10}$ related contours have overlapped with the EM contours where the directions are changed.
The shapes of contours $\delta_{0}-w_\mathrm{{eff0}}$  are narrow for all the constraints as well.

Surprisingly, in the constraining results of SS\Romannum{3}$_{HT1}$  data, the best fitted result of  $\Omega_{m0}$ reaches as large as $0.435^{+0.019+0.037}_{-0.019-0.037}$ which are out of all the existing reasonable constraints. 
There is a  tension between the EM and SS\Romannum{3}$_{HT1}$ data.
And by comparing the results from SS\Romannum{1}$_{\Lambda}$  and SS\Romannum{2}$_{HT10}$,   the model effect brings a  slight shift of the best fitted value of $ \Omega_{m0}$ which is $\Delta \Omega_{m0}=0.004$ in  $f(Q)_{HT1}$ model. While the extra friction term  brings a rather large shift of the best fitted value of $\Omega_{m0}$ which is $\Delta \Omega_{m0}=0.090$ by comparing SS\Romannum{2}$_{HT10}$ and SS\Romannum{3}$_{HT1}$ constraints.

The $\delta=0$ and $w_\mathrm{{eff0}}=-1$ are excluded in $1\sigma$ confidence interval in the EM and SS\Romannum{3}$_{HT1}$ results, while it is included in the SS\Romannum{1}$_{\Lambda}$ related results. 
The evolutions of $\delta$ cross $0$ and the evolutions of $w_\mathrm{{eff}}$ cross $-1$ in most $1\sigma$ intervals. And with increasing of $z$, $\delta$s gradually approach $0$, $w_\mathrm{{eff}}$s gradually deviate from $-1$.
These results are consistent with the analytic calculations in Eqs.(\ref{w2}) and (\ref{deltah}).

As Table \ref{tab1} shows, the $\chi^{2}_{PantheonPlus}$,  $\chi^{2}_{CC}$ and  $\chi^{2}_{DESI}$  of $f(Q)_{HT1}$ model are smaller than that of $\Lambda$CDM model. As a result, $\Delta AIC_{EM}=-14.8$ and  $\Delta BIC_{EM}= -9.3$ which denote the $f(Q)_{HT1}$ model is favored by the EM data. And in the SS\Romannum{1}$_\Lambda$ constraint, $\Delta AIC_{SS\Romannum{1}_\Lambda}=3.2$ which denotes the $f(Q)_{HT1}$ model is favored and $\Delta BIC_{SS\Romannum{1}_\Lambda}=8.1$ which denotes the $f(Q)_{HT1}$ model is ``punished". Considering the large value of $\Omega_{m0}$ in the  SS\Romannum{3}$_{HT1}$ constraint, the future standard siren data may rule out the $f(Q)_{HT1}$ model.

\subsection{Discussion on the $f(Q)_{HT2}$ model}
As Fig. \ref{fht2} shows, the evolution values of $D_L$ related to SS\Romannum{3}$_{HT2}$ simulation  are much smaller than that of SS\Romannum{2}$_{HT20}$  and SS\Romannum{1}$_{\Lambda}$ simulation.
And compared with the  PantheonPlus data, the simulated  SS\Romannum{3}$_{HT2}$ data, which are based on a large negative $\delta_{0}=-0.415$, have much smaller $D_L$ values.

The EM constraint is the tightest one as well. But when the SS\Romannum{3}$_{HT2}$ data is used, it could not give out an effective $\Omega_{m0}$ even after set a prior  $0<\Omega_{m0}<0.4$ as Table \ref{tab} and Fig. \ref{fht2} show. 
Explicitly,  the constraining results of SS\Romannum{3}$_{HT2}$ is $\Omega_{m0}=0.398^{+0.002+0.002}_{-0.001-0.004}$ under the prior $0<\Omega_{m0}<0.4$ with $\chi^2_{SS\Romannum{3}_{HT2}}=1197.2$. Because of such poor fitting results, we do not plot the evolutions of $\delta$ and $w_\mathrm{{eff}}$ for this model.

As Table \ref{tab1} shows, the $\chi^{2}_{CC}$ and $\chi^{2}_{DESI}$ of $f(Q)_{HT2}$ model are smaller than that of $\Lambda$CDM model. While, the $\chi^{2}_{PantheonPlus}$ of $f(Q)_{HT2}$ model is larger than that of $\Lambda$CDM model. Furthermore, $\Delta AIC_{EM}=6.3$ and $\Delta BIC_{EM}=11.8$ which denotes the $f(Q)_{HT2}$ model is excluded by the EM data.
And, based on SS\Romannum{3}$_{HT2}$ simulation and constraint, we conclude that the $f(Q)_{HT2}$ model will be ruled out by the future standard siren observational data as well.

\subsection{Short summary on the non $\Lambda$CDM-like models}
Here, the model effects are much smaller than the extra friction  effects in both models. While the two effects are comparable in the $\Lambda$CDM-like models, e.g.the model effect of $f(Q)_{PE}$ model ($\Delta\Omega_{m0}=0.003$) is $12.5\%$ of the $1\sigma$ range of EM constrained $\Omega_{m0}$ ($\Delta\Omega_{m0}^{1\sigma}=0.024$). As for the extra friction  effect of $f(Q)_{PE}$ model ($\Delta\Omega_{m0}=0.065$), it  is $270.8\%$ of the $1\sigma$ range of EM constrained $\Omega_{m0}$. As the errors caused by model effect could be at the level of $10\%$, it should not be ignored.

And, the Hubble tensions in the non $\Lambda$CDM-like models are slightly larger than that in the $\Lambda$CDM-like models. Anyway, compared with the $D_L$s derived from $f(Q)_P$ and $f(Q)_E$ models, the ones related to  the simulated SS\Romannum{3} data of non $\Lambda$CDM-like model are smaller. Especially, $D^{SS\Romannum{3}_{HT2}}_L$s   seem to smaller than the PantheonPlus data.
All the $\Omega^{SS\Romannum{3}}_{m0}$ for the non $\Lambda$CDM-like models are larger than $0.370$ in $2\sigma$ ranges which are out of most existing constraints.
The $f(Q)_{PE}$ model could be ruled out by the EM data, and both the $f(Q)_{HT}$ models will be excluded by the future standard siren data.

\section{Conclusion}\label{summary}
To study the model and extra frictional effects in standard siren simulation, we simulated standard siren data based on $\Lambda$CDM (SS\Romannum{1}$_\Lambda$), based on the $f(Q)$ models with $\delta=0$ (SS\Romannum{2}) and based on the true $f(Q)$ models with $\delta\neq0$ (SS\Romannum{3}) by using the real  EM observational data as baseline. And  two $\Lambda$CDM-like models ( $f(Q)_P$ and $f(Q)_E$) and two non $\Lambda$CDM-like models ($f(Q)_{PE}$ and $f(Q)_{HT}$) are chosen to constrain. 
The  evolution values of $D_L$ related to SS\Romannum{2} and SS\Romannum{3} simulation are $D_L^{SS\Romannum{2}_{PE}}<  D_L^{SS\Romannum{2}_{P}}\simeq D_L^{SS\Romannum{2}_{E}}\simeq D_L^{SS\Romannum{2}_{HT2}}\simeq D_L^{ SS\Romannum{1}_{\Lambda}}< D_L^{SS\Romannum{2}_{HT1}}$ and $D_L^{SS\Romannum{3}_{HT2}}\ll  D_L^{SS\Romannum{3}_{HT1}}\simeq D_L^{SS\Romannum{3}_{PE}}<D_L^{SS\Romannum{3}_{P}}< D_L^{SS\Romannum{3}_{E}}\simeq D_L^{ SS\Romannum{1}_{\Lambda}}$.

And the tightest constraints are from  the EM data in all $f(Q)$ cosmologies.
The model effects are smaller than  the extra friction  effects in non $\Lambda$CDM-like models. While they are comparable in $\Lambda$CDM-like models. Both effects play important roles in standard siren simulation and could not be ignored.
By comparing the constraining  results, especially the $\chi^2$, AIC and BIC, the $f(Q)_P$ and $f(Q)_E$ models need more observational data (e.g.growth factor) to further study; the $f(Q)_{PE}$ model could be ruled out by the EM data; and both the $f(Q)_{HT}$ models will be excluded by the future standard siren data.
\section{Acknowledgments}
YZ is supported by National Natural Science Foundation of China under Grant No.12275037 and 12275106. DZH is supported by the Talent Introduction Program of Chongqing University of Posts and Telecommunications (grant No. E012
A2021209), the Youth Science and technology research project of Chongqing Education Committee (Grant No. KJQN20230
0609).

\bibliographystyle{spphys}
\bibliography{fq}

\end{document}